\documentclass{revtex4}
\usepackage{amssymb}
\usepackage{amsmath}
\usepackage{graphicx}

\newcommand{\be}{\begin{equation}}
\newcommand{\ee}{\end{equation}}
\newcommand{\ket}[1]{\ensuremath{\left| {#1} \right>}}
\newcommand{\bra}[1]{\ensuremath{\left< {#1} \right|}}

\newcommand{\create}{\ensuremath{\,\hat{a}^{\dagger}}}
\newcommand{\destroy}{\ensuremath{\,\hat{a}}}

\newcommand{\bi}[1]{\ensuremath{\boldsymbol #1}}
\newcommand{\Ca}[1]{\ensuremath{^{\rm #1}{\rm Ca}^+}}

\begin{document}




\title{Quantum science and metrology with mixed-species ion chains}

\author{J. P. Home}

\affiliation{ Institute for Quantum Electronics, ETH Zurich, Switzerland}

\begin{abstract}
 This chapter reviews recent developments in the use of mixed-species ion chains in quantum information science, frequency metrology and spectroscopy. A growing number of experiments have demonstrated new methods in this area, opening up new possibilities for quantum state generation, quantum control of previously inaccessible ions, and the ability to maintain quantum control over extended periods. I describe these methods, providing details of the techniques which are required in order to work with such systems. In addition, I present perspectives on possible future uses of quantum logic spectroscopy techniques, which have the potential to extend precision control of atomic ions to a large range of atomic and molecular species.
\end{abstract}

\maketitle

%
%
%

\section{Introduction}
This review provides describes recent work on the control and manipulation of the quantum states of trapped atomic ions in ion chains containing more than one atomic species. This work has built upon the high precision which has been achieved in the control of the quantum states of trapped atomic ions over the past thirty years (relevant reviews include \cite{08Blatt, 03Leibfried2, 98Wineland2}), but has required a number of new techniques in order to deal with multiple masses of ion in a chain, and provided new opportunities for extending control to new ion species.

The cooling of one type of ion by a second ion of a different type stored in the same ion trap was first observed by \cite{80Drullinger} using multiple isotopes of magnesium, and subsequently  by \cite{86Larson}, who demonstrated cooling of beryllium ions by laser cooling co-trapped mercury ions. Both of these early experiments were performed in Penning traps (\cite{BkGhosh}). This review will focus on more recent work in Paul traps, typically involving small numbers of stored ions (between 2 and 4). The recent revival of interest in such systems has been motivated by two developments. Firstly, work on scalable quantum information processing requires the ability to cool the motion of trapped ions close to the quantum ground state while maintaining quantum superpositions of their internal states. Direct laser cooling involves dissipation of the internal states of the ions, which would destroy the stored information. Co-trapping of a second species of ion allows laser cooling to be performed, where the first species is ``sympathetically'' cooled due to the Coulomb repulsion between ions. The second development stems from the proposal by David Wineland (\cite{98Wineland2}) that in addition to the ability to sympathetically cool ion species that could not be directly Doppler cooled,  the multi-qubit gate techniques developed for trapped-ion quantum information processing could allow state preparation and readout of such species. This opens up the possibility to transfer high precision quantum control techniques to a wide range of atomic and molecular ions.

This review follows the following format. Section \ref{sec:TrapModes} provides an introduction to the mass-dependent pseudopotential and the resulting normal modes of the ion chain in the limit that the excursions of the ions about their equilibrium positions are much smaller than the inter-ion distances. This forms the basis for all experiments described later in the paper. Section \ref{sec:cooling} provides a description of the effect that the form of these normal modes has on the cooling dynamics. Since the ion order affects the dynamics of the ion motion in the normal modes, it is often necessary to initialize a particular ion order. Two techniques for doing this are outlined in section   \ref{sec:reordering}. These sections form the basis for all of the later parts of the review.

The final sections describe applications of the control of multi-species ion chains. Section \ref{sec:quantumlogicreadout} describes the use of one species to read out another, while section \ref{sec:quantumcomputation} describes the use of sympathetic cooling in quantum information processing experiments. Two applications to quantum state engineering are described in section \ref{sec:quantumstateengineering}. Section \ref{sec:molecules} describes challenges and proposals for extending quantum control techniques to molecular ions, including the use of sympathetic cooling and quantum logic techniques to prepare and measure these systems. The final section concludes.

\section{Normal modes of mixed-species chains} \label{sec:TrapModes}

\subsection{Mass-dependent potentials and stability}
The experimental work considered in this review primarily concerns collections of small numbers of ions in radio-frequency Paul traps. A detailed account of the motion of single ions in these potentials can be found elsewhere (examples include \cite{98Wineland2, 03Leibfried2, 69Dehmelt, BkGhosh}). The trapping potential is created using a combination of static and radio-frequency electric potentials applied to the trap electrodes. In this case the total potential experienced by the ion can be written as
\begin{equation}
\phi(\bi{r}_{\rm tot},t) = \phi_{\rm st}(\bi{r}_{\rm tot}) + \phi_{\rm RF}(\bi{r}_{\rm tot}) \cos(\Omega_{\rm RF}t) \ , \label{eq:pot}
\end{equation}
where $\bi{r}_{\rm tot}$ is the position vector of the ion. Under conditions which are generally applicable in the work described below, the motion of the ion can be split into two components, $\bi{r}_{\rm tot} = \bi{r} + \bi{r}_\mu$ where the second component accounts for driven ``micro-motion'' at the drive frequency $\Omega_{\rm RF}$, and the dynamics of the secular motion with co-ordinate $\bi{r}$ can be approximated by solving the equation of motion of the particle of charge $q$ and mass $m$ in a time-averaged pseudopotential (\cite{69Dehmelt})
\be
\Phi(\bi{r},m) = \phi_{\rm st}(\bi{r}) + \Phi_{\rm pond}(\bi{r}, m) \ , \label{eq:pseudopotential}
\ee
with the ponderomotive part of the potential  given by a static term
\be
\Phi_{\rm pond}(\bi{r}, m) = \frac{q |\nabla \phi_{\rm RF}|^2}{4 m \Omega_{\rm RF}^2 } \ .
\ee
This expression is obtained by averaging the kinetic energy of the driven micro-motion of the ion at frequency $\Omega_{\rm RF}$. The secular motion of strings of ions in the pseudopotential approximation will be our principal concern in the remainder of this review. The validity of the pseudopotential approximation is discussed in depth by \cite{69Dehmelt} and \cite{BkGhosh}.

\subsection{Stability of ion motion}
The trapped-ion experiments considered in this review use potentials which can be approximated close to the trap center ($x = y = z = 0$) using a harmonic expansion
\begin{eqnarray}
\phi_{\rm st}(\bi{r}) &=& \frac{V}{\rho^2} \left(\chi_x x^2 + \chi_y y^2 + \chi_z z^2\right) + \bi{E}\cdot\bi{r}, \nonumber \\
\phi_{\rm RF}(\bi{r}) &=& \frac{U_{\rm r.f.}}{\rho^2} \left(\kappa_x x^2 + \kappa_y y^2 + \kappa_z z^2 \right) \cos(\Omega_{\rm RF}t)\ \ . \label{eq:phiRF}
\end{eqnarray}
 where $\bi{r} \equiv (x,y,z)$, $\rho$ is a characteristic length scale for the trap (e.g. the distance from the ion to the nearest trap electrode) and $\chi_{x,y,z}$, $\kappa_{x,y,z}$ are dimensionless parameters which depend on the trap geometry and the applied voltages. $\bi{E}$ is a static electric field. In order to satisfy Laplace's equation at all times we require that $\chi_x + \chi_y + \chi_z = 0$ and $\kappa_x + \kappa_y + \kappa_z = 0$. Experimentally the electric field component can often be controlled independently of the potential curvature, and is usually tuned to zero. This results in the minimum of both the static and pseudopotential terms co-existing at the origin of the potential. While it is common in experimental work to apply a single voltage amplitude $U_{\rm r.f.}$ for the radio-frequency component of the total potential, in many recent experiments the static part of the potential is produced by several potentials applied to multiple independent electrodes. In these cases $V$ is not a single value. Nevertheless, the form given above can be used, with $V$ giving an order of magnitude estimate for the voltages, and the details of the potential produced by the exact configuration absorbed into the $\chi_{x,y,z}$.

For the oscillating potential given in equation \ref{eq:phiRF}, the ponderomotive part of the potential can be found to be
\be
\Phi_{\rm pond}(\bi{r}, m) = \frac{q U_{\rm r.f.}^2}{m \Omega_{\rm RF}^2 \rho^4}\left(\kappa_x^2 x^2 + \kappa_y^2 y^2 + \kappa_z^2 z^2 \right) \ \ .
\ee
and the secular oscillation frequencies of the ion in the total pseudopotential are given by
\be
\omega_{x,y,z} = \sqrt{\frac{2 q^2 U_{\rm r.f.}^2 \kappa_{x,y,z}^2}{ m^2 \Omega_{\rm RF}^2 \rho^4} + \frac{2 q V \chi_{x,y,z}}{m \rho^2}}
\ee
The mass dependence of the ponderomotive potential is of much importance in working with mixed-species ion crystals. In terms of the stability of ion motion, care should be taken to ensure that the ion motion is stable for \emph{both} ion species.  In the potential given above, stability of a single ion can generally be obtained for secular frequencies $\omega_{x,y,z} \leq \Omega_{\rm RF}/(2 \sqrt{2})$. A rough guideline for achieving stable motion in mixed-species chains is that this stability criterion should be satisfied for single ions of both ion species independently. It should be noted that for larger numbers of ions it is often found that the secular frequencies of motion should be below the limit given here due to anharmonic effects stemming from the Coulomb interaction. More precise discussions of the stability of ion motion can be found from solutions to the Matthieu equations which can be formed from the equations of motion; discussions are given in earlier reviews, for example \cite{03Leibfried2}.

\subsection{Normal mode analysis} \label{sec:normalmodes}
In this section we review the general approach to calculating the equilibrium positions and motional vibrations of a set of ions in a single potential well such as that described above. For the purpose of this description, we assume the ions to be at low enough temperature that the oscillations about their equilibrium positions are small compared to the inter-ion separations. The derivation follows the methods of \cite{98James1}, extending these to mixed-species chains as was done by \cite{00Kielpinski} and \cite{01Morigi}. \cite{00Kielpinski} studied motion in all three spatial dimensions for chains consisting of odd numbers of ions, with a single ion of a different species to the rest placed at the center of the chain. \cite{01Morigi} investigated the motional modes and cooling dynamics of the axial motion of ion strings containing two species, focused on systems of two and three ions.

The energy of a string of $N$ ions in the potential given in equation \ref{eq:pseudopotential} is given by $E = T + U$ where
\begin{eqnarray}
	T & = & \sum_{j=1}^{N}\frac{m_{j}}{2}\dot{\bi{r}}_{j}^{2},\nonumber \\
	U & = & \sum_{j=1}^{N}q \Phi(\bi{r}_{j},m_j)+\frac{1}{2}\sum_{{j,l=1\atop j\neq l}}^{N}\frac{q^{2}}{4\pi\epsilon_{0}|\bi{r}_{l}-\bi{r}_{j}|}\ \ \ .
\end{eqnarray}
Here $m_{j},\bi{r}_{j}$ ($\bi{r}_l$) denote the mass and position of the $j$th ($l$th) ion, and $q$ is the charge (the ions are here assumed to have the same charge). In what follows, it is convenient to write the coordinates as $3N$ scalar parameters $z_{1},\ldots,z_{3N}$ denoted by the subscript $i$, that produce the $N$ vectors $\bi{r}_{1},\ldots,\bi{r}_{N}$ with subscript index $j$.

Solving the set of equations $\partial U/\partial z_{i}=0$ gives the set of equilibrium positions $\{z_{i}^{0}\}$ for the ions. In practice for longer chains numerical methods must be used, in which case the direct approach is to apply a minimization routine to the potential energy $U$.

In a Taylor expansion of the potential around
the equilibrium positions, the leading term is
at second order, which gives the $3N \times 3N$ symmetric Hessian matrix
\begin{equation}
	 H_{ik}^{\prime}=\left.\frac{1}{\sqrt{m_{i}m_{k}}}\frac{\partial^{2}U}{\partial z_{i}\partial z_{k}}\right|_{\{z_{i}^{0}\}} .
	\label{eq:hessian}
\end{equation}
At this point it is convenient to switch to mass-weighted coordinates (see for example \cite{LLmechanics}) $z_{i}^{\prime}=\sqrt{m_{i}}z_{i}$, allowing the kinetic energy $T$ to be written in a form that is independent of mass. This substitution is the primary difference between the treatments of normal modes of homogeneous ion chains (e.g. \cite{98James1}) and the mixed-species case which we consider here.

The normal modes and their
corresponding frequencies can be found by solving the Lagrangian
equations of motion using the displacements
from equilibrium, $\zeta_{i}^\prime=z_i^\prime-z_{i}^{0\prime}$. Neglecting higher orders in $U$ than those described by equation \ref{eq:hessian}, the $3N$ equations of motion are
\begin{equation}
	 \ddot{\zeta}_{k}^{\prime}+\sum_{i=1}^{3N}H_{ik}^{\prime}\zeta_{i}^{\prime}=0 .
	 \label{eq:vector_transformation_into_normal_mode_basis}
\end{equation}
Inserting a fiducial solution $\zeta_k^\prime = \zeta_k^{0\prime}e^{i\omega t}$ gives a linear system of equations that can be diagonalized to yield the normal modes of the system, which are defined by the eigenvalues
and eigenvectors of the matrix $H_{ik}^{\prime}$. The eigenvalues
are equal to $\omega_{\alpha}^{2}$, where $\omega_{\alpha}$ is the
motional frequency of the normal mode $\alpha$. The matrix of eigenvectors
$e'_{i,\alpha}$ allows the normal-mode coordinates to be expressed as
a function of the individual ion coordinates by use of
\begin{equation}
	 \zeta_{\alpha}^{\prime}=\sum_{i=1}^{3N}e'_{i,\alpha}\zeta_{i}^{\prime}.
\end{equation}
and since the Hessian is a real symmetric matrix, we can also write the inverse relation
\be
    \zeta_{i}^{\prime}=\sum_{\alpha=1}^{3N}e'_{i,\alpha}\zeta_{\alpha}^{\prime}.
\ee
Since each normal mode acts as an independent oscillator, they can be
quantized in the usual manner, by writing the mass-weighted position and momentum operators as
\begin{eqnarray}
	 \hat{\zeta}_{\alpha}^{\prime}&=&\frac{\sigma_{\alpha}^{\prime}}{\sqrt{2}}\left(\hat{a}_{\alpha}+\hat{a}_{\alpha}^{\dagger}\right) ,
	\label{eq:quantized_position_operator} \\
\hat{p}'_\alpha &=& i \frac{\hbar}{\sqrt{2} \sigma_{\alpha}^{\prime}}\left(\hat{a}_{\alpha}-\hat{a}_{\alpha}^{\dagger}\right) \ \ . \label{eq:palpha}
\end{eqnarray}
 where $\sigma_{\alpha}^{\prime} = \sqrt{\hbar/\omega_{\alpha}}$ and $\hat{a}_\alpha^\dagger$, $\hat{a}_\alpha$ are the raising and lowering ladder operators. The quantized form for the $i$th ion's excursion from equilibrium is then given by (\cite{01Morigi})
\begin{equation}
\hat{\zeta}_{i}=\frac{1}{\sqrt{ m_{i}}}\sum_{\alpha=1}^{3N} e'_{i,\alpha} \sigma_{\alpha}^{\prime}\left(\hat{a}_{\alpha}+\hat{a}_{\alpha}^{\dagger}\right) .
\end{equation}
The root-mean-square position uncertainty of the motional component $\zeta_i$ in the normal mode denoted by $\alpha$ is therefore given by $\sqrt{\hbar/(2 m \omega_\alpha)} e_{i,\alpha}$ when the mode is in the ground state.

\begin{table}
\centering
\begin{tabular}{|c|rrr|rrr|} \hline & \multicolumn{3}{|c|}{$^{9}$Be$^+$ ion} & \multicolumn{3}{|c|}{$^{24}$Mg$^+$ ion} \\ \hline
$\omega_\alpha/(2 \pi)$ & $e'_{x_1,\alpha}$ & $e'_{y_1,\alpha}$ & $e'_{z_1,\alpha}$ & $e'_{x_2,\alpha}$ & $e'_{y_2,\alpha}$ & $e'_{z_2,\alpha}$  \\
\hline
12.11 &  1.000  & 0 & 0 & 0.018 & 0 & 0 \\
11.03 & 0  & 1.000 & 0  & 0  & 0.020 &  0 \\
4.68  &  0.018       & 0    & 0   & -1.000    & 0 & 0  \\
4.04 &  0 & 0 & -0.926 & 0 & 0 & 0.378 \\
3.53 &  0 & 0.020 & 0 & 0 & -1.000 & 0\\
1.90 &  0 & 0 & 0.378 & 0 & 0 & 0.926 \\
\hline
\end{tabular}
\caption{Normal mode eigenfrequencies and eigenvectors for a beryllium-magnesium ion chain in a trap for which a single beryllium ion has secular frequencies of $[\omega_x, \omega_y, \omega_z] = 2 \pi \times ([12.26, 11.19, 2.69]~{\rm MHz})$, and magnesium has secular frequencies $[\omega_x, \omega_y, \omega_z] = 2 \pi \times ([3.72, 4.82, 1.65]~{\rm MHz})$. The root-mean-square position uncertainty of the motional component $\zeta_i$ in the normal mode denoted by $\alpha$ is $\sqrt{\hbar/(2 m \omega_\alpha)} e_{i,\alpha}$ when the mode is in the ground state. The index $i$ takes values $x_1, y_1, z_1$ for the beryllium ion, and $x_2, y_2, z_2$ for the magnesium ion. Values of zero are quoted at the same level as precision as the non-zero values, with zeros after the decimal place removed for ease of viewing.}
\label{tab:normalmodes}
\end{table}

\subsection{Normal mode characteristics}

The standard setup in which the experiments described in the rest of this review have been performed involve a pseudopotential which is zero along one axis. If this axis is denoted $z$, we therefore have that $\kappa_x = -\kappa_y$ and $\kappa_z = 0$. The static potential is then arranged to provide confinement along the $z$ direction which is weaker than the confinement of any of the ions in the $x-y$ plane. The ion chain then lines up along the axis of the trap. For small numbers of ions in such a potential, analytical expressions for the eigenmodes can be derived, however even for two ions the mass difference adds significant complexity. Explicit expressions for a pair of ions in standard potentials of the form described above are given by \cite{12Wubbena}.

The normal modes have several characteristic features, which depend on the relative masses of the two ion species, and on the relative contributions of the ponderomotive potential and the static potential to the confinement of the ions. Modes of oscillation of the ion chain involve different combinations of ions which oscillate either in phase with each other or out-of-phase.  An example case for $^{9}$Be$^+$ and $^{24}$Mg$^+$ which illustrates the most important features is given in table \ref{tab:normalmodes}. From this example it can be seen that while the axial modes ($z$)  involve significant motion of all of the ions for both the in-phase and out-of-phase motion, the radial modes ($x$, $y$) of the two species are almost independent (one ion has $e'_{i,\alpha}$ close to zero, with that of the other ion close to one). This structure can be observed in the normal modes of many mixed-species ion chains. The independence of the radial modes is due to the difference in radial confining potentials for the two different ion species, and can be understood by considering the motion of each individual ion as an independent oscillator. These oscillators are coupled together by the Coulomb interaction, which has a strength which depends on the equilibrium spatial separation. In the radial directions, if the ions were not coupled they would have very different oscillation frequencies. In the coupled case, this means that the exchange of energy due to the Coulomb interaction is far from resonance and has a small effect on the resulting motion. This effect gets more pronounced as the mass ratio of the two ion species increases.

\subsection{Shifts in normal modes due to additional effects}
Though most experiments performed to date make use of the standard linear trap potential described in the previous section, in practice imperfections mean that this is a simplification of the real potential. A number of factors which have no impact on the characteristics of the normal modes of single-species ion chains do have an impact on the normal modes of mixed-species chains. These result from terms which have a different effect on each ion species. Three examples are given in the following list.

\begin{enumerate}
  \item A pseudopotential gradient along the ion chain axis $z$ will produce a differential force on ions of different mass. Such a gradient can arise as part of trap fabrication imperfections or for trap designs which deviate significantly from the ideal linear Paul trap (\cite{10Amini}). A pseudopotential gradient exerts a force of $P = -q \nabla{\Phi}_{\rm pond}$ on an ion of mass $m$, and a force of $P/\mu$ on  a second ion of mass $\mu \times m$. This will change the equilibrium spacing of the ions by approximately $\pm P(1 - 1/\mu)/(m \omega^2)$ where $\omega$ is the single ion oscillation frequency for the ion of mass $m$. The sign of this change will depend on the order of the ion crystal, with the shorter separation resulting from an order in which the pseudopotential force points from light to heavy ion. This in turn produces a shift in the normal mode frequencies. For the axial mode of a two ion crystal and values of $\mu$ which are close to one, the primary effect is to shift the frequency of the out-of-phase mode, since this has a significant contribution from the Coulomb interaction, the strength of which depends on the inter-ion distance. \cite{11Home} observed the effect of a pseudopotential gradient of $\sim 0.2$~eV/m by looking at the shift in the axial modes of a $^{24}$Mg$^+$--$^{9}$Be$^+$ ion pair as a function of ion order. The out-of-phase mode of motion shifted by 2.5~kHz between the $^{24}$Mg$^+$--$^{9}$Be$^+$ and $^{9}$Be$^+$--$^{24}$Mg$^+$ ion orderings.

  \item A second source of differential forces on the two ions is the radiation pressure force due to laser cooling light used for Doppler cooling, which will only affect the ion which interacts with the radiation.  The effect of this force can be estimated by assuming a two level ion driven by a laser tuned close to resonance with a dipole allowed transition. The time-averaged force on the ion is
\be
\bi{F} = \hbar \bi{k} \Gamma \rho_{ee}
\ee
where $\bi{k}$ is the wavevector of the light, $\Gamma$ is the natural linewidth of the atomic transition and $\rho_{ee} = \bra{e}\hat{\rho}\ket{e}$ is the steady-state probability of finding the ion in the excited state of the transition, given by
\be
\rho_{ee} = \frac{s/2}{1 + s + (2 \Delta/\Gamma)^2} \label{eq:rhoee}
\ee
with $s = 2|\Omega|^2/\Gamma^2$. $\Omega = \mu_{\rm eg}E/\hbar$ is the Rabi frequency for a light field with peak electric field $E$ interacting with a transition with dipole moment $\mu_{\rm eg}$ and $\Delta$ is the detuning of this light field from resonance (more details can be found in standard textbooks, for example \cite{BkMetcalf}).  If directed along the axis of a two-ion crystal where only one of the ions is illuminated this force will result in an change of the equilibrium spacing of two ions of $\epsilon = |\bi{F}|/(m \omega^2)$. The effect on two equal mass ions (with only one illuminated by the light), is to modify the frequency of the out-of-phase motional mode along the axis of the trap by a factor $1 \pm \epsilon/(2 d)$ where $d$ is the ion separation in the absence of the force. This gives the order of magnitude of frequency shifts which could be expected. As a simple example, consider a pair of equal mass calcium ions for which only one is cooled using a laser at 397~nm. If $\Delta = -\Gamma/2$ and $s = 1$, the magnitude of the force is $F = 3.4\times 10^{-20}$~N, resulting in a relative displacement of $\epsilon \simeq 13$~nm for a trap in which the axial oscillation frequency of a single calcium ion is 1~MHz. The shift in the out-of-phase mode frequency is then around 2~kHz. The change in ion separation $\epsilon$ is similar to the root-mean-square ground state wavefunction size of $11$~nm for a single trapped calcium ion in the same potential.

\item The third factor which can lead to deviation of the normal modes of mixed-species ion chains is anharmonicity in the trapping potentials, which was studied theoretically and experimentally by \cite{11Home}. Odd-order anharmonicities produce shifts which are dependent on the ion order for asymmetric ion chains containing ions of different mass. For the experiments in \cite{11Home}, shifts of 20~kHz were observed between the magnesium-beryllium and beryllium-magnesium ion orders for a trap in which a single beryllium ion had an axial secular frequency of $\omega_s = 2 \pi \times (2.7~{\rm MHz})$. If this shift is ascribed only to a cubic term in the potential, this would correspond to a term $(m_{\rm Be} \omega_s^2 z^2/2) (z/\lambda_3)$ with $\lambda_3 = 230$~$\mu$m (\cite{11Home}).

All of the effects described above can be measured if re-ordering techniques are available for an asymmetric mixed-species ion chain consisting of two unequal mass ions (see section \ref{sec:reordering}). The relevant information can be obtained by differences in normal mode frequencies between the two configurations $(m_1, m_2)$ and $(m_2, m_1)$, where $m_1, m_2$ label the two ions, and the ion order should be read from left to right. This was used by \cite{11Home} in order to characterize both the anharmonicity of a trap and the pseudopotential gradient. For measuring the effect of laser intensity an alternative would be to change the intensity of the laser while monitoring the motional frequencies.
\end{enumerate}

\subsection{Normal mode diagnostics}\label{sec:modediagnostics}
Frequency and amplitude characteristics of normal modes can be extracted by a number of techniques. These are not unique to mixed-species work, and thus the reader is referred elsewhere for an in depth discussion, for example see \cite{98Wineland2}. The primary methods used involve coherent transitions on resolved motional sidebands of narrow-linewidth transitions. To illustrate these methods, we can consider a single travelling wave optical field which is tuned close to the resonance frequency $\omega_0$ of a transition between two internal states $\ket{g}$ and $\ket{e}$. We assume that the laser interacts only with ion $j$ with co-ordinate $\bi{r}_j$, and that the lifetime of the two levels is long compared to the timescales of interest.  Working in the interaction picture with respect to the Hamiltonian $H_0 = (\hbar \omega_0/2) \left(\ket{e}\bra{e} - \ket{g}\bra{g} \right)$, and making a rotating wave approximation with respect to the optical frequencies, the interaction between the laser and the ion can be written as
\be
\hat{H}_I = \frac{\Omega}{2} \exp({i \bi{k}\cdot\hat{\bi{r}}_j}) e^{i \delta t} \hat{\sigma}_+ + {\rm h.c.}
\ee
where $\Omega$ is the Rabi frequency with which coherent oscillations between the two states involved would be driven in the limit that the ion mass tends to infinity, $\hat{\sigma}_+ = \ket{e}\bra{g}$ is the raising operator for the two internal states involved, and $\delta = \omega_L - \omega_0$ is the detuning of the optical field from the internal state transition. The Rabi frequency $\Omega = \mu_{\rm eg} E_0/\hbar$, where $\mu_{\rm eg}$ is the transition dipole moment and $E_0$ is the zero-peak amplitude of the electric field of the light (\cite{BkMetcalf}). The phase of the laser has been neglected here, since it is not of importance for any of the techniques described below. The $\exp({i \bi{k}\cdot\hat{\bi{r}}_j})$ factor is the operator which describes the atomic recoil when a single photon is absorbed (see e.g. \cite{79Wineland}). The exponent can be written in terms of the normal mode co-ordinates as
\be
i {\bi{k}\cdot\hat{\bi{r}}_j} =  i \sum_\alpha \eta_{j, \alpha} \left(\create_\alpha  + \destroy_\alpha \right) \ ,
\ee
where we have defined the Lamb-Dicke parameter for the oscillation of ion $j$ in the normal mode $\alpha$ as
\be
\eta_{j, \alpha} =  \sqrt{\frac{\hbar}{2 m_j \omega_\alpha}} \bi{k}\cdot\bi{e}'_{j,\alpha} \label{eq:LambDicke}
\ee
and $\bi{e}'_{j,\alpha} = (e'_{j_x,\alpha}, e'_{j_y,\alpha},e'_{j_z,\alpha} )$ is the vector describing the components of motion of ion $j$ in the motional mode $\alpha$ (The integers $j_x, j_y, j_z$ indicate the values of the index $i$ which correspond to the $x$, $y$ and $z$ components of motion of ion $j$). The Lamb-Dicke parameter squared gives the ratio $E_{R, \alpha}/(\hbar \omega_\alpha)$, where $E_{R, \alpha}$ is the recoil energy transferred to the mode $\alpha$ when a  single photon is absorbed along the laser beam direction.

It is useful to move to the interaction picture with respect to the Hamiltonian describing the energies of the normal modes
\be
\hat{H}_m = \sum_\alpha \hbar \omega_\alpha \left(\create_\alpha \destroy_\alpha + 1/2 \right) \ ,
\ee
which results in an interaction picture Hamiltonian for the ion-laser coupling of the form
\be
\hat{H}_I = \frac{\Omega}{2} \exp\left[i \sum_\alpha \eta_{j, \alpha} \left(\create_\alpha e^{-i \omega_\alpha t} + \destroy_\alpha e^{i \omega_\alpha t}\right)  \right] e^{i \delta t} \hat{\sigma}_+ + {\rm h.c.} \ \ .
\ee

Spectroscopy of the normal modes is performed by scanning the detuning $\delta$ over a frequency range encompassing the modes in question. For $\delta = +\omega_\alpha$, the resonant terms in the Hamiltonian can be found up to first order of the expansion of the exponential in $\eta_{j, \alpha}$ to be
\be
\hat{H}_{+} = \frac{\Omega}{2}  i \eta_{j, \alpha} \create_\alpha  \sigma_+ + {\rm h.c.}
\ee
which results in Rabi oscillations between the states $\ket{g, n_\alpha}$ and $\ket{e, n_\alpha + 1}$ with a Rabi frequency proportional to $\sqrt{n_\alpha + 1}$ (\cite{98Wineland2}). This is often referred to as a blue-sideband transition. In what follows, we denote this type of transition by $R_+(\theta)$, with $\theta = \Omega \eta_{j, \alpha}\sqrt{n_\alpha + 1} t_p$ and $t$ the duration for which the laser is applied. For laser pulse durations $t_p$ such that $\theta = \pi$, the internal state makes a transition, and the motional state is changed by a single quantum. For $\delta = -\omega_\alpha$, a similar Hamiltonian is derived, but with the positions of  $\create_\alpha$ and $\destroy_\alpha$ exchanged. This results in Rabi oscillations between the states $\ket{g, n_\alpha}$ and $\ket{e, n_\alpha - 1}$ with a Rabi frequency proportional to $\sqrt{n_\alpha}$  (referred to as a red-sideband transition). This will be denoted by $R_-(\theta)$, with $\theta = \Omega \eta_{j, \alpha}\sqrt{n_\alpha} t_p$.  In order to make measurements of normal mode frequencies, the laser frequency is scanned and the relative positions of the resonances on $R_+$ and $R_-$ are recorded (in the ideal case, the difference frequency between these two equals twice the mode frequency). For sensitive measurements of motional frequencies of normal modes, it is worth being aware of off resonant processes, which can produce AC Stark shifts of the transitions involved (\cite{97Steane2}). An alternative approach to coherent manipulations of internal and motional states which has been used in much of the work with mixed-species ion crystals involves the use of Raman transitions. These can be described by similar equations to those used above, but with the laser frequency $\omega_L$ and wavevector $\bi{k}$ replaced by the difference frequency and difference wavevector of the two Raman laser beams (see for example \cite{98Wineland2}), which are given by $\delta\omega$ and $\delta\bi{k}$ respectively.

An alternative method for measuring normal mode frequencies which is free of AC Stark shifts can be obtained using oscillating electric fields applied to a nearby trap electrode. In the limit that the ion chain is much shorter than the ion-electrode distances, the field can be approximated as being the same at each ion. In this case, the Hamiltonian describing the motional coupling of the field $\bi{E}(t)$ to mode $\alpha$ is
\be
\hat{H}_{\bi{E}, \alpha}(t) =  q \bi{E}(t)\cdot \left(\sum_j \bi{e}'_{j,\alpha}/ \sqrt{m_j} \right) \hat{\zeta}'_\alpha \label{eq:EfieldHam} \ \ .
\ee
For modes which couple to the electric field (for which $\bi{E}(t)\cdot \left(\sum_j \bi{e}'_{j,\alpha}/ \sqrt{m_j} \right) \neq 0)$, the oscillating field results in a displacement of the ion motional state which is enhanced when the field is resonant with the motional mode. For an ion initialized in the ground state prior to application of the field, this results in a coherent state of motion. The excitation of the ion due to this displacement operation can be detected by a number of methods. The most sensitive of these is to prepare the ion close to the ground state prior to application of the electric field, and to use laser-driven motional sideband transitions between internal states to detect any motion afterwards. The Rabi frequencies on the motional sidebands are proportional to $\sqrt{n_\alpha}$ and $\sqrt{n_\alpha + 1}$ for $R_-$ and $R_+$ respectively, thus a Fourier decomposition of a sideband Rabi oscillation can be used to extract motional occupations (\cite{03Leibfried2}). For detecting excitation from the ground state, the motion-subtracting sideband is particularly useful. If the ion is in the ground state prior to the sideband pulse, no transition can occur because this requires a quantum of motion to be removed. If a transition does occur, this indicates that the motion must have started in an excited state. This method is therefore sensitive to excitation at the single quantum level. Note that depending on whether the ion is initialized in the higher or lower internal state, either $R_-$ or $R_+$ will subtract motional quanta.

The use of a detuning $\delta = 0$ can also be used to detect motional excitation. The carrier transition between states $\ket{e, n_\alpha}$ and $\ket{g, n_\alpha}$ has a Rabi frequency which can be written to second order in the Lamb-Dicke parameter as
\be
\Omega_c(n_\alpha) \simeq \frac{\Omega}{2} \left( 1 - \frac{1}{2}\sum_\alpha \eta_{j, \alpha}^2 (2 n_\alpha + 1) \right) \ .
\ee
Due to the quadratic dependence on $\eta_{j, \alpha}$, this method is most effective for moderate values of the Lamb-Dicke parameter, and is less sensitive to motional excitations than driving a motional sideband. However it is more robust to changes in the AC Stark shifts of the internal state transition, because the transition is power broadened. In \cite{11Home}, the carrier transition was used to detect the excitation of the axial in-phase mode of motion of a beryllium-magnesium ion pair with a Lamb-Dicke parameter of 0.18. In this case the Rabi frequency of the $n = 17$ state is roughly half that of the ground state.

Resonant excitation of a motional mode can also be detected without cooling the mode to the motional ground state, though in general these methods are not as sensitive as those described above. Large motional excitations can produce Doppler shifts which approach the linewidth of a dipole transition, resulting in a reduction in the observed ion fluorescence signal (\cite{07Wesenberg}). For a single ion, motional excitations up to vibrational quantum numbers of around 1000 are required to obtain a significant signal. Though this approach is commonly used for single ions, it does not work as well for diagnosing modes of longer chains, primarily due to the anharmonicity of the Coulomb interaction between the ions, which causes the motional frequencies of the ions to shift before significant motional excitation can be induced.

If a laser is tuned to the side of the resonance profile of an allowed transition, the scattering rate of the photons will be modulated in phase with the coherent motion of the ion. By recording the time of arrival of the scattered photons and correlating this with the phase of the motional excitation, high sensitivity to coherent motion can be achieved. This has been demonstrated by \cite{09Vahala, 10Biercuk} and used for detection of excitations with mixed-species ion chains by \cite{11Hume}.

\subsection{Compensation of stray fields}
\begin{figure}
\centering
\includegraphics[width=0.8\textwidth]{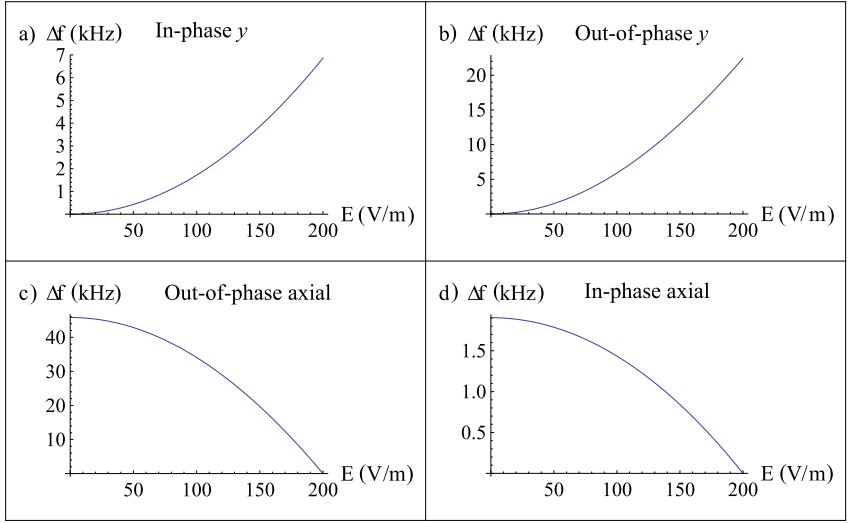}
\caption{Calculated frequency shifts of the normal modes of a magnesium-beryllium ion pair as a function of the radial electric field strength. The electric field is applied parallel to the $y$ axis of a trap for which a single beryllium ion would have trap frequencies of $f_z = 2.7~{\rm MHz}$,$f_x = 12.26 ~{\rm MHz}$, $f_y = 11.18~{\rm MHz}$. The modes shown exhibit motion along $z$ and $y$ for zero field ( modes exhibiting motion along the $x$ direction have a negligible field shift for the sizes of fields used in this figure). The plots are a) the in-phase radial mode directed along $y$, b) the out-of-phase mode directed along $y$, c) the axial out-of-phase mode, and d) the axial in-phase mode of motion. } \label{fig:fieldshifts}
\end{figure}

The mass dependence of the ponderomotive part of the pseudopotential means that by contrast with single-species ion chains, the normal modes of mixed-species chains change in the presence of additional static fields. The heavier mass ions experience weaker confinement from the ponderomotive term, so a radial static field will displace the heavy ions from the pseudopotential null line by more than those of the lighter ion species. This distortion of the equilibrium positions of the ions in the crystal leads to changes in the frequencies and eigenvectors of the normal modes of the ion chain. As an example,  figure \ref{fig:fieldshifts} plots the calculated frequency shifts as a function of a radial field along the $y$ axis for a beryllium magnesium ion pair in a trap with single beryllium ion secular frequencies $f_z = 2.7~{\rm MHz}$, $f_x = 12.26 ~{\rm MHz}$, $f_y = 11.18~{\rm MHz}$. The mode amplitudes for a radial field of 200~Vm$^{-1}$ are shown in table \ref{tab:normalmodesfield}.

\begin{table}
\centering
\begin{tabular}{|c|rrr|rrr|} \hline & \multicolumn{3}{|c|}{$^{9}$Be$^+$ ion} & \multicolumn{3}{|c|}{$^{24}$Mg$^+$ ion} \\ \hline
$\omega_\alpha/(2 \pi)$ & $e'_{x,\alpha}$ & $e'_{y,\alpha}$ & $e'_{z,\alpha}$ & $e'_{x,\alpha}$ & $e'_{y,\alpha}$ & $e'_{z,\alpha}$ \\ \hline
 12.11 &   1.000 & 0 & 0 & 0.018 & 0 & 0 \\
 11.06 & 0 & -0.999 & 0.024 & 0 & -0.016 & -0.014 \\
 4.67  &  0.018 & 0 & 0 & -1.000 & 0 & 0 \\
 4.04  &  0 & -0.017 & -0.817 & 0 & -0.470 & 0.334      \\
 3.42  &  0 & -0.027 & -0.450 & 0 & 0.882 & 0.137          \\
 1.89  & 0 & -0.005 & 0.360 & 0 & 0.038 & 0.932 \\
\hline
\end{tabular}
\caption{Normal mode eigenfrequencies and eigenvectors for a beryllium-magnesium ion chain under the same conditions as those used for table \ref{tab:normalmodes} but with an additional field of 200~Vm$^{-1}$ applied along the radial $y$ direction.}
\label{tab:normalmodesfield}
\end{table}

Since radial stray fields lead to ions sitting at positions where the oscillating component in the electric field is not zero, the micromotion of these ions is increased. This can cause a variety of problems for precision spectroscopy and quantum control of trapped ions. For precision spectroscopy micromotion leads to Doppler shifts of transition frequencies which are difficult to constrain (\cite{98Berkeland}). For quantum control tasks micromotion leads to an effective modulation of control fields. If the micromotion changes, the modulation index also shifts, leading to a loss of control. For both single-species and multi-species ion chains, micromotion is typically detected by monitoring the effect of the modulation on the ion's interaction with light fields, either using Doppler cooling or narrow-linewidth transitions (\cite{98Berkeland2, ThRoos}). The normal mode shifts and changes in the crystal equilibrium positions described above provide new tools for compensating these stray fields. This was first realized by \cite{03Barrett}, who proposed two methods for doing this which are outlined below.

The first method involves measuring the frequencies of the normal modes, which shift as a function of the applied field. This method requires a good understanding of the normal modes and of the other parameters which form the trapping potential, and is therefore a challenging method for more than a few ions. If the mode frequencies are well understood, changes in frequency as a function of a given applied field can then be identified. It can be seen from the example given in figure \ref{fig:fieldshifts} b) that by tuning the field such as to minimize the mode frequency of the out-of-phase $y$ mode, stray fields in the $y$ direction can be reduced to zero. This method is limited by the resolution with which the mode frequencies can be determined, and the sensitivity of the mode frequencies to electric fields. The frequency uncertainties quoted by \cite{11Home} are of order 200~Hz, which would allow stray fields of below 15~V/m to be detected in this trap. Higher sensitivity to stray fields could be produced by reducing the pseudopotential confinement of the trap.

A second method which is simple to implement with Raman transitions involves lining up the axis of the ion crystal with the difference vector between the wavevectors of the two Raman beams. If the difference wavevector $\Delta \bi{k}$ is aligned along the axis of the trap, the Raman beams should only couple to motional modes which involve motion along the axis of the trap. When no stray radial fields are present, only the axial modes have this property. In the presence of stray fields, modes which principally exhibit motion perpendicular to the axis have components along the trap axis, and thus couple to the Raman light. It is therefore possible to align the direction of the normal modes with $\Delta \bi{k}$ by adjusting the radial electric field such that the Rabi frequency of the radial sidebands is minimized relative to the axial modes. The alignment of $\Delta \bi{k}$ with the trap axis can be verified  by minimizing the coupling of the Raman light to radial modes using a single ion.

\section{Sympathetic cooling} \label{sec:cooling}
One important use of mixed-species ion chains is to use accessible cooling transitions in one ``refrigerator'' ion species to cool another ``target'' species. This allows the internal states of the target to remain unaffected while the cooling of the external degree of freedom proceeds. It also allows target ion species to be cooled without requiring the ability to laser cool these species directly.

\subsection{Laser cooling of mixed-ion chains}

For cooling in both the Doppler and resolved sideband limits, cooling relies on the coupling of the ion motion to an optical field via the internal states (\cite{79Wineland}). To understand Doppler cooling of an ion chain, we consider a simplified picture of an ion with two internal states which are  coupled by a laser light field close to resonance with a dipole-allowed transition. The transition linewidth is assumed to be much larger than the motional frequencies, ie. $\Gamma >> \omega_\alpha$ for all motional modes. We will assume that only ion $j$ is cooled. The rate at which the kinetic energy of the ion $j$ is cooled can then be found using a semiclassical treatment (\cite{BkMetcalf}) to be
\be
\frac{d E_{\rm K. E.}^{j}}{dt} = \frac{2 \hbar  \Gamma}{m_j}\left(\frac{d \rho_{ee}}{d\Delta}\right)\frac{\left(\bi{k}\cdot \hat{\bi{p}}_j\right)^2}{2 m_j}  \label{dEkindt}
\ee
with $\rho_{ee}$ as defined in equation \ref{eq:rhoee}. The momentum component $\hat{p}_i$ of the ion corresponding to the $i$th co-ordinate can be expressed in terms of the normal mode co-ordinates as
\be
\hat{p}_i = - i \hbar \frac{\partial}{\partial \zeta_i} = -i \hbar \sum_{\alpha = 1}^{3 N} \frac{\partial \zeta'_\alpha}{\partial \zeta_i} \frac{\partial}{\partial \zeta'_\alpha} =  \sum_{\alpha=1}^{3N} \sqrt{m_i} e'_{i,\alpha} \hat{p}'_\alpha \label{eq:pi}
\ee
which allows us to write
\be
\bi{k}\cdot \hat{\bi{p}}_j = \sum_{\alpha=1}^{3N} \sqrt{m_i} \bi{k}\cdot\bi{e}'_{j,\alpha}  \hat{p}'_\alpha \label{eq:kpj} \ ,
\ee
where again we have used $\bi{e}'_{j,\alpha} = (e'_{j_x,\alpha}, e'_{j_y,\alpha},e'_{j_z,\alpha} )$.

Substitution of equations \ref{eq:palpha}, \ref{eq:LambDicke} and \ref{eq:kpj} into equation \ref{dEkindt} results in a cooling rate of the mode in question which can be expressed in terms of quanta per second as
\be
\frac{d n_\alpha}{dt} =  2 \omega_\alpha \eta_{i, \alpha}^2 \left(n_\alpha + 1/2\right) \Gamma \left(\frac{d \rho_{ee}}{d\Delta}\right) \ \ . \label{eq:coolrate}
\ee
The cooling rate thus depends quadratically on the Lamb-Dicke parameter $\eta_{j, \alpha}$ for the cooling ion $j$ in mode $\alpha$. In addition to the frictional cooling, the random nature of absorption and spontaneous emission events heat the mode $\alpha$ through momentum diffusion (\cite{BkMetcalf}). In the limit that $\Omega << \Gamma, \Delta$, each random emission/absorbtion event can be assumed to be independent, and the heating is well approximated by
\be
\frac{d n_\alpha}{dt} = \left(\eta_{j, \alpha}^2 + \frac{(2/5) \hbar^2 |\bi{k}|^2 |\bi{e}'_{j, \alpha}|^2}{2 m} \right) \Gamma \rho_{ee} \ \ . \label{eq:heatrate}
\ee
The first term in this expression describes heating due to the absorbtion of photons from the cooling laser, while the second is the contribution from recoil of the ion during spontaneous emission, which is an average over all directions of the emitted photons assuming a dipole emission pattern. By comparing equations \ref{eq:coolrate} and \ref{eq:heatrate} we see that both heating and cooling processes depend quadratically on the  magnitude of the ion motion in the normal mode $|\bi{e}'_{j,\alpha}|$. Thus if these are the only heating and cooling processes present we attain the usual Doppler limit temperature, which is independent of $|\bi{e}'_{j,\alpha}|$ (see for example \cite{03Leibfried2}).

Similar arguments hold for cooling in the resolved sideband limit. For continuous sideband cooling in the Lamb-Dicke regime (for which $\eta_{j, \alpha}^2 (2 n_\alpha + 1) \ll 1$), the rate of cooling of a given mode is proportional to $\eta_{j,\alpha}^2$. The cooling limit does not depend on the Lamb-Dicke parameter if only off-resonant transitions contribute to heating (\cite{03Leibfried}).

In real systems, additional heating mechanisms need to be overcome (see next section). Therefore it is desirable that high cooling rates can be achieved. We can see from equation \ref{eq:coolrate} that in order to perform effective cooling of a particular shared mode of motion a significant Lamb-Dicke parameter is required for the ion which is being actively cooled. For \emph{radial} modes of mixed-species ion chains with different ion masses, this poses a problem. As the mass ratio between the two ions increases, $|\bi{e}'_{j,\alpha}|$ is rapidly reduced to close to zero for one of the ion species, and tends to 1 for the other (this can be seen in table \ref{tab:normalmodes} for a two-ion crystal with a mass ratio of $24/9$). In view of the mode amplitude imbalances, cooling a target species $X$ is best performed using a refrigerant ion with a similar mass. The sympathetic cooling efficiency of a spectroscopy ion by an ion of a different species is considered in detail by \cite{12Wubbena}, who consider effects such as the relative masses and the dependence on the static and ponderomotive components of the pseudopotential $\Phi(\bi{r}, m)$.

\subsection{Heating due to fluctuating electric fields}
 Ions stored in a trap experience fluctuating electric fields which lead to heating of the motional modes. Known sources of fluctuations are due to noise on supply voltages and Johnson noise due to resistances in lines connecting the trap to the voltage source. It is generally possible to suppress these sources through careful design of electronics and by placing filters between the trap electrode and the source. Calculations of the effect of Johnson noise predict electric field fluctuations which are orders of magnitude below those observed in many trapped-ion experiments. The amount of heating is therefore ``anomalous'', due to a process which is not well understood. Recent work on electrode surface cleaning indicates that this is due to contaminants on the surface on the electrodes (\cite{12Hite, 11Allcock}), which produce scaling of electric field noise which is consistent with models of large numbers of microscopic fluctuating patch charges or dipoles (\cite{00Turchette, 06Deslauriers, 11Danilidis, 11Safavi}). The fluctuations are found to be reduced by cooling the trap (\cite{06Deslauriers, 08Labaziewicz}). The heating rate for trapped ions due to this anomalous heating scales roughly as $d^{-4}$ where $d$ denotes the ion-electrode distance, thus these problems become more severe in the small-scale traps built for multiplexed ion trap quantum information processing (see \cite{10Amini} for a review).

The relation between the fluctuations in the electric field and the heating rate of normal modes can be derived from equation \ref{eq:EfieldHam}, which describes the coupling of each normal mode to a uniform field. Making the assumption that the fluctuations of the field are stationary noise processes with correlation times short compared to the change in state of the ion, the rate of heating from the ground state of motion can be calculated to be
\be
\dot{n}_\alpha = \frac{q^2}{4 \hbar \omega_\alpha} S_{E, \alpha}
\ee
where
\be
S_{E, \alpha} = 4 \int_0^{\infty} \xi(t) \xi(t +  \tau) e^{i\omega_\alpha \tau}d\tau
\ee
with $\xi(t) = \bi{E}(t)\cdot\left(\sum_j \bi{e}'_{j,\alpha}/\sqrt{m_j}\right)$. $S_{E, \alpha}$ reduces to $S_{E}(\omega_\alpha)/m$ for a single ion of mass $m$ and an electric field noise spectral density of $S_E(\omega_\alpha)$ (\cite{00Turchette}). For an ion crystal consisting of multiple ions of a single species, the centre-of-mass mode has the largest coupling to the electric field, and antisymmetric modes with $\xi(t) = 0$ are not heated by uniform electric fields. In multi-species ion crystals, ion configurations with symmetry about the center of the ion chain do have purely antisymmetric modes, but other configurations do not tend to exhibit these properties.

\subsection{Improving cooling rates for radial modes}
The laser cooling rates of the radial modes can be increased by adding radial static field components  to the standard trapping potential (\cite{10HumeThesis}). Since the primary radial confinement is due to the mass-dependent ponderomotive potential, these result in the heavier ion being pushed off the axis of the trap, so that the vector connecting the equilibrium positions of the two ions is no longer aligned with the zero of the pseudopotential. Table \ref{tab:normalmodesfield} gives the calculated mode eigenvector components with a static electric field of $200$~Vm$^{-1}$ applied in the radial $y$ direction and all other trap parameters set to the same values used for table \ref{tab:normalmodes}. For the low frequency radial normal modes there is now significant motion for \emph{both} ion species, meaning that either ion could cool these modes at a significant rate. By contrast, the higher frequency radial modes still have a much lower amplitude for the heavy ion, which could not be used to cool these ions with a high rate. Note that this method will also involve significant micromotion of the heavy ion, which may be undesirable if it is being used in high-precision spectroscopy.

Obtaining significant motion of the heavy ion in the high frequency radial modes is challenging due to the large frequency difference between these radial modes and the axial modes, which means that any coupling between the ions' motion in the two directions is far from resonance. Increasing the axial confinement is in general undesirable because it leads to a point where the ion chain makes a transition from a linear chain to a zig-zag configuration (\cite{97Steane2}), at which point some of the low frequency radial modes (which primarily involve motion of the heavier ion species) are close to zero frequency. If cooling \emph{all} modes of the chain is necessary, it is desirable to use two species of ion which are similar in mass. If this is not an option, then the lighter mass ion should be used as the coolant. This is because laser cooling of the lighter species can be used to cool the highest radial modes, and the low-frequency radial modes can be mixed with axial modes using additional radial fields. Experimental studies of sympathetic cooling and quantum control which have been performed to date used mass ratios between 1/3 to 25/9, where the numerator represents the cooled ion species.

\section{Re-ordering ions of different mass} \label{sec:reordering}
The normal mode frequencies and amplitudes of mixed-species chains depend on where in the chain the heavier mass ions are situated. A two-species linear ion chain consisting of $N$ ions of which $k$ are ions of one species has $N!/(k!(N - k)!)$ possible ion configurations. Since the normal modes depend on the ion order, it is highly desirable to be able to initialize a single configuration, or a subset of the possible ion orderings. Once a particular configuration has been initialized, it will stay fixed unless the ions gain enough energy that the ion chain disintegrates (laser cooling is then usually used to dissipate this energy in order that the ion chain re-forms into a crystal). This can happen through collisions with high energy background gas particles, which typically occur once every few minutes in a room-temperature ultra-high-vacuum environment. The correct ion order can be verified by looking at fluorescence (for ions where this is observable) on a camera with sufficient spatial resolution to separate individual ions in the chain. Where this is not possible, (such as where fluorescence detection is not sensitive to ion position), an alternative method is to perform spectroscopy of the ion motion. This can be performed using the methods described in section \ref{sec:modediagnostics}.

The ion re-ordering techniques which have been used in experiments with mixed-species ion chains to date make use of the mass-dependence of the total potential. These techniques were first proposed and developed by Till Rosenband and co-workers for use in the mixed-species ion chain quantum logic spectroscopy experiments described in section \ref{sec:quantumlogicreadout}. They have also been used extensively in the NIST quantum information experiments which are described in section \ref{sec:quantumcomputation}. In what follows I give two specific examples from the latter, which illustrate the methods involved. It should be noted that deterministic re-ordering of ions of the same species has been performed by control of the trapping potentials in a single trap zone by \cite{09Splatt}, and using junctions in trap arrays (\cite{06Hensinger, 09Blakestad}). These methods would work for mixed-species ion chains, but perform two-directional swapping of ion positions. This contrasts with the methods described below, which deterministically achieve a chosen ion order from any given starting configuration.

\subsection{Symmetric ordering, heavy ions centered}
In the experiments of \cite{09Jost}, a symmetric ion order was desired with two heavier mass magnesium ions ($^{24}$Mg$^+$) placed between two of a lighter species ($^9$Be$^+$). The re-ordering method in this case made use of the contrast between the curvature of the radial potential, which is primarily provided by the mass-dependent pseudopotential, and that of the axial potential which is dominated by static terms. Increasing the axial curvature has two effects on the curvature of the total potential (trap plus Coulomb repulsion) experienced by each ion. It produces anti-confining static terms in the radial trap potential, and also pushes the ions in the chain closer together, which results in a larger radial anti-confinement due to the mutual Coulomb repulsion of the ions. Both of these effects reduce the radial curvature of the total potential experienced by each ion. When the radial curvature goes to zero for one of the ions the ion chain starts to deform into a configuration for which some ions have an equilibrium position away from the zero of the pseudopotential. This point will depend on the relative masses and positions of the ions in the chain, but since the heavier ions are more weakly confined by the pseudopotential, in many cases these are the first to move off-axis. For a given starting ion configuration, the initial off-axis state is metastable, ie. one of the other ion orders results in a lower energy equilibrium configuration. As the axial confinement is increased further, a critical point is reached at which the off-axis ions are radially displaced by a large enough distance that they can access the genuine lowest energy configuration, independent of the starting ion order. Since ions on the end of the chain are separated by larger distances than in the center, ion configurations with the heavy ions on the ends of the chain tend to require larger changes in trap parameters in order to reach the critical point at which re-ordering will occur.

\begin{figure}
\centering
\includegraphics[width=0.8\textwidth]{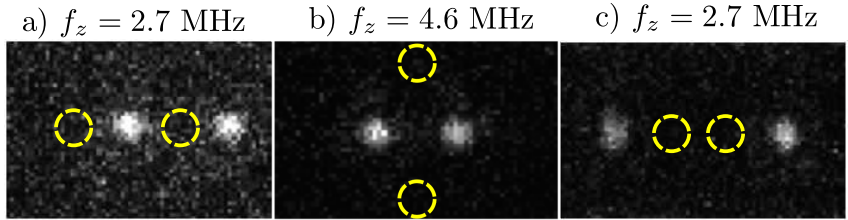}
\caption{Trap frequencies and ion configurations used in the 3-step re-ordering of ion chains consisting of two beryllium and two magnesium ions performed in the NIST group for the experiments in \cite{09Jost}. The pictures are images taken on a CCD camera during the re-ordering. The expected positions of the magnesium ions are indicated by dashed circles. The frequencies given are for a single beryllium ion. Radial frequencies are given in the main text. Picture taken by John Jost, NIST 2008.} \label{fig:4ionreorder}
\end{figure}

The beryllium and magnesium ions used by \cite{09Jost} have a mass ratio of $M_{\rm Be}/M_{\rm Mg} = 9/24$. The potential well in which experiments were being performed on linear chains of ions had an axial curvature leading to a trap frequency for a single beryllium ion of 2.7~MHz, with radial frequencies of 12.2~MHz and 11.2~MHz (this corresponds to an axial frequency of 1.65~MHz for a single magnesium ion, with radial frequencies of 4.8~MHz and 3.7~MHz). To re-order the ions, the voltages on the control electrodes of the trap were ramped over $\sim1$~ms to a potential in which a single beryllium ion has an axial frequency of 4.6~MHz, and radial frequencies of 9.7~MHz and 12.9~MHz. Taking these frequencies and calculating the expected magnesium secular frequencies gives 1.5~MHz and 5.4~MHz in the radial directions, and 2.8~MHz in the axial direction. With these parameters the four ion crystal with two magnesium and two beryllium ions finds the lowest energy for the ``diamond'' configuration shown in figure \ref{fig:4ionreorder}b), and will reach this configuration from any starting order of the ions. The trap voltages are then ramped back to their original values, with the ions relaxing back to a configuration in which the heavy ions are centered. Both during the transition to the diamond configuration and the relaxation into the final linear chain, the ions gain significant amounts of energy, which must be extracted by laser cooling in order to reach the temperatures required for quantum control.

\subsection{Asymmetric ordering}
\begin{figure}
\centering
\includegraphics[width=0.8\textwidth]{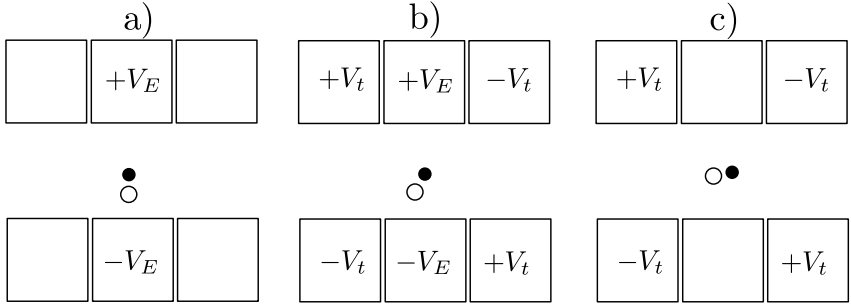}
\caption{Schematic showing the shifts in potentials and ion positions used for re-ordering a two-ion multi-species ion chain. Each square represents an electrode to which static potentials can be applied. The resulting ion configurations are shown using a filled circle for the light ion and an outline for the heavy ion. Step 1 a) First a radial field is applied with sufficient strength to push the ion pair into a radial crystal. Step 2 b) An additional static ``twist'' trapping potential is applied which provides a quadrupole potential aligned at a non-zero angle with the pseudopotential axis of the trap. This breaks the symmetry, pushing the heavy ion to the left of the light ion. Step 3 c) The field is removed, resulting in a near axial crystal. In the final step, the additional ``twist'' potential is removed, and the ions return to the trap axis.} \label{fig:2ionasymmetric}
\end{figure}

As an alternative to the symmetric ordering described above, it may be desirable to place one species of ion to one side of the ion crystal. This has been done in a number of experiments with two ion crystals (\cite{10HumeThesis, 10Chou, 11Home}). Here I describe the method used to re-order beryllium and magnesium ions by \cite{11Home}. The first step involves applying a static field in a direction perpendicular to the $z$ axis, which is defined by the pseudopotential null (the ion chain is initially aligned with this axis). The field displaces the heavy ion radially by more than the light ion, due to the relative strengths of the radial pseudopotential confinement. Above a critical value of the field, this results in the ion chain making a sudden transition to a lowest energy configuration in which the ion chain is aligned along the additional field (shown in figure \ref{fig:2ionasymmetric} a)). For starting conditions for which a single beryllium ion has an axial frequency of 2.7~MHz and radial frequencies 12.2~MHz and 11.2~MHz, the required field for radial alignment of a beryllium-magnesium ion chain is 900~V/m along the radial direction of weakest confinement. Removing the radial field at this point would not result in any particular desired order, since there is symmetry along the axis of the trap. An extra step is therefore required to break the symmetry. In the experiments of \cite{11Home}, the symmetry breaking was achieved by twisting the axis of the static component of the trapping potential, using potentials of the form shown in figure \ref{fig:2ionasymmetric} b) applied to the segmented electrodes of the trap. This displaced the heavy ion relative to the light ion along the $z$-axis . While this ``twist'' potential set was applied, the bias field applied in the first step was removed, bringing the heavy ion back close to the trap axis, but now in the desired position (figure \ref{fig:2ionasymmetric} c)). The final step removes the ``twist'' potential, resulting in an axial ion chain with the desired ion order. This method should be extendable to longer chains of ions. Depending on the relative masses and the size of the electric fields and ``twist'' potentials, alternating configurations or configurations with all heavy ions to one side of all light ions may be possible. Alternative methods to break the symmetry include any methods for applying differential forces to the ions, including either applying a pseudopotential gradient or the use of radiation pressure.

\section{Quantum-logic readout}\label{sec:quantumlogicreadout}
The ability to trap and manipulate two ion species in a single ion trap opens up the control of trapped ions to species beyond those which have accessible Doppler cooling transitions. These include many of the atoms in the periodic table, as well as the molecular species considered in section \ref{sec:molecules}. In order to perform more advanced control of any of these species, a pre-requisite is the ability to perform spectroscopy on them. An important application of high precision spectroscopy of ions is in atomic frequency standards work, where work with mixed-species chains of trapped ions is currently state of the art ( see for example \cite{10Chou}).

Two problems occur in trying to control ion species with no closed transition manifolds on which Doppler cooling can be performed. The first is in trying to cool the ion. Aside from laser cooling, no other methods are currently able to cool atomic species from room temperature down to the milli-Kelvin range. Secondly, the closed transition presents a means of extracting information from the system, because the repeated emission of photons provides information about the internal state of the ion. For ions with closed cooling transitions, state-dependent resonance fluorescence is the most common method of measuring the final state of the system. The use of a second ion species can be useful in both of these tasks. It provides a route to cooling complex atoms and ions through the sympathetic cooling of the shared normal modes of motion which was described earlier in this review. It also provides a means to extract information from the system using the technique of Quantum Logic Spectroscopy (QLS), which has been proposed and pioneered by Dave Wineland's research group at NIST (\cite{98Wineland2, 05Schmidt}).

Quantum logic spectroscopy makes use of the common motion of the two trapped ions to transfer information from the spectroscopy ion (which does not have a closed cooling transition) to a cooling/readout ion of a species which can be fully controlled by standard techniques of ion trap control. The mapping of information from the spectroscopy ion to the motion can be performed using a number of different methods, which are described below in order of their first experimental demonstration.

\begin{figure}
\centering
\includegraphics[width=0.8\textwidth]{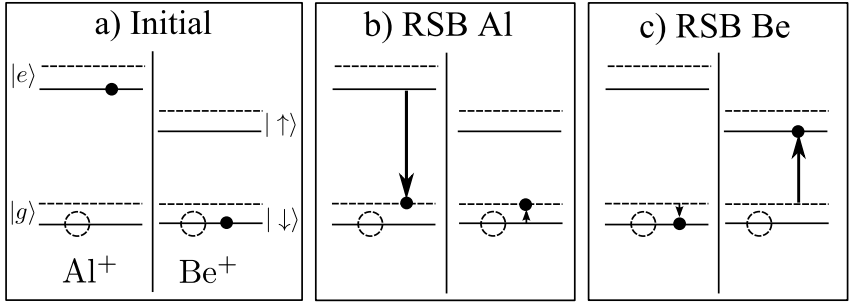}
\caption{Sketch of the quantum logic scheme demonstrated by \cite{05Schmidt}. The internal state levels of Al$^+$ are labeled as $\ket{g}= \ket{^1{\rm S}_0,F = 5/2, M_F} $, $\ket{e} = \ket{^3{\rm P}_1, F' = 7/2, M_{F}'}$. For beryllium the levels are $\ket{\uparrow} = \ket{^{2}{S}_{1/2}, F = 1, M_F = -1} $, $\ket{\downarrow} = \ket{^{2}{S}_{1/2}, F = 1, M_F = -2}$ . The motional energy level of each ion is indicated by a solid line ($n = 0$) and a dashed line for $n = 1$. At each step, the final ion states of the beryllium and aluminium ions are marked with a solid or unfilled disk depending on the starting condition. The thick arrow marks the transition in the ion which is driven using a laser, while the thin arrow corresponds to the resulting transition of the motional state in the other ion.} \label{fig:QLS1}
\end{figure}

Quantum logic spectroscopy was first demonstrated by \cite{05Schmidt}, who performed readout of the internal state of an aluminium ion using a co-trapped beryllium ion. The sequence of steps required to do this is illustrated in figure \ref{fig:QLS1}. The measurement protocol was used to determine whether an aluminium ion was in the $\ket{g} \equiv \ket{^1{\rm S}_0, F = 5/2, M_F}$ state or the $\ket{e} \equiv \ket{^3{\rm P}_1, F' = 7/2, M_{F}'}$ state (the values of $M_F$, $M_F'$ are left undefined here because the researchers demonstrated the technique using a range of values). The measurement protocol starts with the axial modes of the two-ion crystal cooled to the ground state. A motional sideband $R_-(\pi)$ is then driven on the $\ket{g, n_\alpha} \leftrightarrow \ket{e, n_\alpha - 1}$ transition using a laser tuned below the internal state transition by the trap oscillation frequency $\omega_\alpha$. In the ideal case, for an ion in the $\ket{g, n_\alpha = 0}$ state, no transition is driven because this would require the removal of a single quantum of motion, which is not available since the system is already in the ground state. If the aluminium ion starts in the $\ket{e, n_\alpha = 0}$ state, the laser drives a transition $\ket{e, n_\alpha = 0} \rightarrow \ket{g, n_\alpha = 1}$, exciting the collective motion. The remainder of the protocol uses the beryllium ion to read out the motional state of the normal mode. This was performed using a stimulated Raman transition between the $\ket{\downarrow} \equiv \ket{F = 2, M_F = 2}$ and $\ket{\uparrow} \equiv \ket{F' = 1, M_F' = 1}$ hyperfine states of the beryllium readout ion, with the Raman difference frequency and pulse durations tuned to the motion subtracting sideband transition $\ket{\downarrow, n_\alpha} \leftrightarrow \ket{\uparrow, n_\alpha-1}$. The beryllium ion starts in the lower energy $\ket{\downarrow}$ state, having been prepared in this state during the ground state cooling. If the motion has been excited by the aluminium transition, the Raman pulse then results in an internal state transition, otherwise no change in internal state occurs. The internal state of the beryllium ion is subsequently read out by state-dependent fluorescence (\cite{98Wineland2}). In the ideal case, the beryllium state is correlated with the original state of the aluminium ion, and thus the protocol implements the desired measurement.

A similar method to the readout described above allowed \cite{05Schmidt} to perform state preparation of a single hyperfine state of the ion. The procedure makes use of repeated mapping of the internal state to the motion, allowing ground state cooling to be used as a dissipative step in an internal state optical pumping process. An ion in the $\ket{^{1}S_0, F = 5/2, M_F}$ state was transferred to another ground state with $M_F' = M_F + 1$ using a two-pulse coherent transfer involving a carrier transition from $\ket{^{1}S_0, F = 5/2, M_F}$ to $\ket{^{3}P_1, F = 5/2, M_F}$ state followed by a coherent transfer to  $\ket{^{1}S_0, F = 5/2, M_F + 1}$ using a motion-adding sideband transition. The coherent transitions are reversible, which does not satisfy the requirements for state preparation (where any initial state must be pumped into a single final state). The critical ingredient for adding dissipation is thus provided by the sideband transition, which adds a single quantum to the motion. This motional excitation is removed by Raman sideband cooling of the ion string using the beryllium ion, providing the required irreversibility. The researchers repeated this irreversible transfer for $M_F = -5/2, -3/2, -1/2, 1/2, 3/2$, resulting in preparation of the ion in the $M_F = + 5/2$ state. Using different sequences of transfer steps, and by implementing the inverse ($M_F \rightarrow M_F - 1$) transfer, each of the hyperfine ground states could be prepared. These methods are natural precursors to the methods described in section \ref{sec:molecules} for rotational state preparation.

The protocols for state preparation and measurement described above are limited in the achievable fidelities by imperfect ground state cooling and inaccuracies in the motional sideband pulses. \cite{05Schmidt} report a success probability lower than 93\% for quantum state readout. The complexity of the transfer process makes it challenging to achieve levels above 99\% for a single-shot readout, as is achieved using state-dependent fluorescence for atomic ions. In many situations in quantum state estimation, the measurement must anyway be repeated many times in order to build up statistics, reducing the requirements on readout fidelity. Higher readout fidelities have been achieved using a modification to the initial protocol, which is described in the next section.

\subsection{Population-preserving ``non-demolition'' readout.}

\begin{figure}
\centering
\includegraphics[width=.8\textwidth]{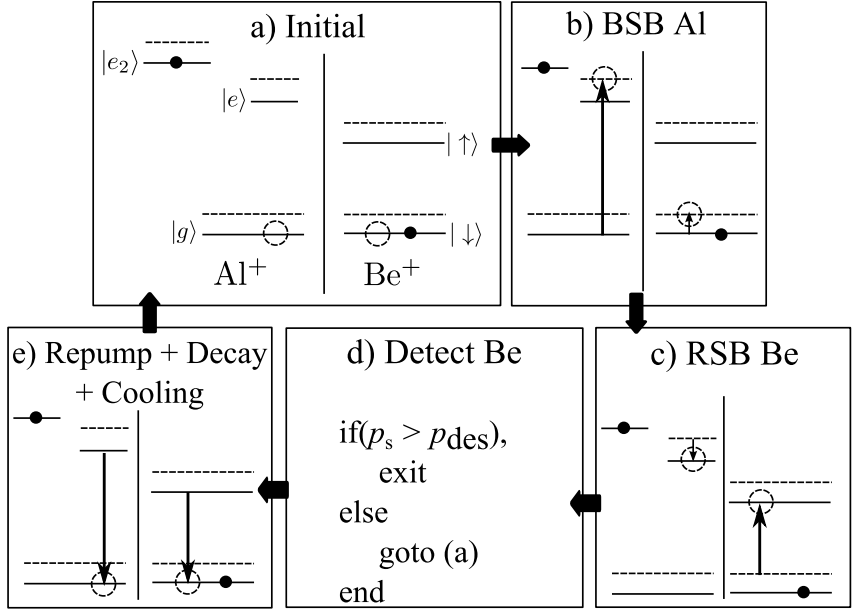}
\caption{Sketch of the quantum ``non-demolition'' logic scheme demonstrated by \cite{07Hume}. Notation is the same as for figure \ref{fig:QLS1}, with the addition of a third level $\ket{e_2} = \ket{^3{\rm P}_0}$ for the aluminium ion. Note that the populations of the aluminium levels at the end of the sequence (e) is the same as at the start (a). This allows the sequence to be repeated, which can be used to improve the fidelity of the measurement.} \label{fig:QLS2}
\end{figure}

Though the original demonstration of quantum logic spectroscopy described in the previous section made use of sideband transitions of the spectroscopy transition itself, these methods were subsequently transferred to readout of the $^{1}$S$_{0} \leftrightarrow ^3$P$_0$ ``clock'' transition in the aluminium ion, which has excellent properties for frequency standards work due to its insensitivity to magnetic and electric field gradients and an extremely small blackbody shift (\cite{07Rosenband, 10Chou}). Though spectroscopy is performed on the clock transition, readout was performed using the $^1$S$_0\leftrightarrow^3$P$_1$ transition, which has a larger matrix element (\cite{07Hume}). The method used differs from that described above, and is illustrated in figure \ref{fig:QLS2}. The ions are initialized in the ground state of motion prior to the readout sequence. A blue sideband is then driven on the $^1$S$_0\leftrightarrow^3$P$_1$ transition. Since only an ion in the $^{1}$S$_{0}$ state will interact resonantly with the light, the sideband transition excites the motion conditional on the internal state. Following this step, the beryllium ion is used to read out the motional state as described in the previous section. The $^3$P$_1$ level has a spontaneous emission lifetime of 300~$\mu$s for decay back to the $^1$S$_0$ level, thus after around a millisecond the aluminium ion returns to the same internal state as was occupied before the measurement started. The population preservation of the aluminium states results in an ideal projective measurement in the energy eigenbasis, which is often referred to as a quantum ``non-demolition'' measurement. It is advantageous because the measurement of the state can be repeated, with the combination of measurement results allowing higher measurement fidelities to be achieved than for a single round. This is also indicated by the sequence shown in figure \ref{fig:QLS2}, which is repeated until the decision point (part d)) exits the loop. The repeated measurement was used by \cite{07Hume} in the following way. If the number of counts recorded at the photo-multiplier tube on performing a measurement of the beryllium ion in measurement $j$ is $n_j$, the probability that this corresponded to the aluminium ion in state $\ket{i}$ can be written as $P(n_j|i)$. Since the detection process is repeated, a set $\{n_j\}$ of photon numbers is recorded. The probability of getting this set of detection results from the state $\ket{i}$ is $P(\{n_j\}|i) = \prod_{j}P(n_j|i)$. The deduced probability to be in the state $i$ is then found from Bayes rule to be
\be
P(i| \{n_j\}) = \frac{P(\{n_j\}|i)}{\sum_k P(\{n_j\}|k)} \ .
\ee
\cite{07Hume} updated this probability after every detection round, and used this to compute the probability of success $p_s = P(i_{\rm max}|\{n_j\})$, where $i_{\rm max}$ is the state with the highest probability given the measurement results. If $p_s$ exceeds a desired threshold $p_{\rm des}$, the detection is halted and the state of the aluminium ion is deduced to be $\ket{i_{\rm max}}$. Otherwise an additional round of measurement is performed. This method was demonstrated experimentally for a measurement sequence which had an accuracy of 85\% for a single round of measurement. Repeated measurement allowed the success probability to be improved by two orders of magnitude to 99.84\%, limited by the rate at which the readout could be repeated relative to the decay lifetime of 21~s of the $^{3}$P$_0$ state (\cite{07Hume}).

\subsection{Readout using coherent state-dependent forces} \label{sec:QLogicCoherent}

The methods described above require resonant excitation of sideband transitions in the spectroscopy ion, and rely on the ability to cool at least one motional mode of the ion chain close to the quantum ground state of motion. In order to make ``non-demolition'' measurements, the spectroscopy ion was also required to spontaneously relax on the sideband transition. To relax these requirements, \cite{11Hume} demonstrated readout using an alternative method of exciting motion in a state-dependent manner. Instead of resonantly driving a sideband transition in the spectroscopy ion, an oscillating state-dependent dipole force was created using a laser detuned from resonance with the $^{1}S_0 \leftrightarrow ^{3}P_1$ transition. The state-dependent force was created using two counter-propagating laser beams with the same polarization but with their relative frequency $\delta \omega = \omega_1 - \omega_2$ tuned close to the oscillation frequency of one of the normal modes of the ion chain. The ion which interacts with the light then experiences a dipole potential of the form
\be
V(z, t) = V_s \cos(\delta{\bf k}\cdot{\bf r} - \delta\omega t) \label{eq:sdforce}
\ee
where $\delta {\bf k}$ is the wavevector difference of light from the two laser beams. $\delta \omega$ was chosen to be close to resonance with the axial in-phase mode of motion of the ion pair, with the result that only this mode is resonantly excited by the optical field. The factor $V_s$ in equation \ref{eq:sdforce} accounts for the dependence of the dipole potential on the internal state of the ion. An expansion of the potential to first order in $\eta_{j,\alpha}$ reveals a term which describes a resonant force on the ion. For larger motional excitation, the higher-order terms in the expansion of the potential become important, limiting the total excitation which can be produced (see experimental work by \cite{07McDonnell} and \cite{10Poschinger} for further details).

\cite{11Hume} used both time-correlated fluorescence detection and resolved Raman sidebands to detect the resulting motional excitation (these methods are described in more detail in section \ref{sec:reordering}). The use of time-correlated fluorescence detection has the advantage that no ground state cooling of the ion is required, but the sensitivity to excitation amplitude is reduced relative to the use of Raman sidebands. The readout method was applied to discriminate between states in the  $^{1}S_0$ and $^{3}P_0$ levels of the aluminium ion, and to discriminate between various $M_F$ states of the $^{1}S_0$ ground level.

The coherent force method has the advantage over resonant sideband excitation that the readout method itself can make use of any method for producing state-dependent forces. In the experiments described above, optical forces were generated using a continuous-wave laser. Additional options would be to generate such a force using static or oscillating magnetic field gradients (\cite{90Dehmelt,11Ospelkaus}), or using pulsed lasers (\cite{12Senko}). These methods widen the potential use of quantum logic spectroscopy techniques to a large range of ion species.

For all quantum logic readout schemes, though entanglement naturally appears during the protocol, there is no requirement in any of the schemes above that quantum coherence should be maintained throughout the protocol. In this regard, the requirements on control are generally relaxed relative to that which is required to perform a two-qubit quantum logic gate.

\section{Quantum computation}\label{sec:quantumcomputation}
One important application of multi-species chains is in scalable methods for trapped-ion quantum information processing. Here pairs of internal states of the ``qubit'' ion species are used for storage of quantum information. A crucial ingredient in the most advanced schemes investigated to date are multi-qubit gates which make use of the shared motion of the ions as a quantum ``bus'' to perform conditional logic. The highest quality multi-qubit gates which have been demonstrated to date use state-dependent forces generated by laser fields to control the ions (\cite{08Benhelm, 09Kirchmair}). The infidelity of a two ion gate performed on two similar mass ions due to a finite Lamb-Dicke parameter has been estimated by \cite{00Sorensen1} to be
\be
1 - F \simeq 0.3 \pi^2 \eta^4 \bar{n}(\bar{n} + 1) \label{eq:gateerror}
\ee
where $\eta$ is the Lamb-Dicke parameter of the driven mode and all modes which couple to the light are assumed to be in thermal equilibrium at the same temperature. This illustrates that motional excitation will reduce the fidelity of multi-qubit gates. For a mixed-species crystal with two logic ions a similar expression would be expected to hold, with each mode which interacts with the laser contributing a similar factor to the infidelity. It is expected that infidelities below 0.0001 will be required to realize fault-tolerant quantum computation (\cite{03Steane}, \cite{05Knill}). Values for $\eta$ used in laser driven multi-qubit logic experiments vary between $0.05$ and $0.25$ (\cite{08Benhelm, 03Leibfried, 05Haljan, 06Home}) requiring motional modes cooled to mean vibration occupations of $1.5$ and $0.01$ quanta respectively to attain ``fault-tolerant'' levels. The use of near-field microwave fields and magnetic field gradients to perform gates provides a much smaller effective Lamb-Dicke factor and hence high fidelity gates would not be expected to be limited by motional excitation even for Doppler cooled ions (\cite{11Ospelkaus}).

In addition to heating of the ions from the field fluctuations described in section \ref{sec:cooling}, motional excitation of ions can also occur due to imperfect implementation of processing steps. To scale up ion trap quantum computation it is necessary to combine techniques for transporting quantum information with high-fidelity gates. The methods which have been demonstrated to date include transporting the ions themselves through an array of traps (\cite{98Wineland2, 02Kielpinski, 02Rowe}), or connecting ions through entanglement with emitted photons (\cite{07Moehring}). Both of these methods lead to some level of motional excitation, which when repeated a large number of times (as is likely to be required for a large-scale quantum computation) would impact on the quality of multi-qubit operations. Deterministic transport of ions through a trap array is performed by applying time-varying control potentials to the trap electrodes. Though in principle this can be performed with no motional excitation, imperfect control of the applied potentials will mean that in general this is not the case. While this does not necessarily result in the ions finishing in a thermal mixture of states, the final state is unknown, and thus a control operation calibrated for ions starting close to a particular state (usually the ground state) would have an increased probability of error. \cite{07Moehring} provided the first demonstration of a photonic quantum information transport protocol between two ions. This method is probabilistic, requiring many repetitions in order to establish entanglement as a quantum information resource between the remote regions of the quantum processor (which can subsequently be utilized for computation using teleportation protocols along the lines proposed by \cite{99Gottesman}). Since each photon emission leads to recoil heating of the motion at the level of $\eta^2$ in units of the trapping frequency, the result of many repetitions will be significant heating, which must be removed if the quantum information stored in the ions is to be further manipulated. In both cases a  cooling mechanism is required which does not affect the stored information; sympathetic cooling satisfies this need.

\subsection{Experimental demonstrations}

Sympathetic cooling of motional modes of ions trapped in Paul traps to the vicinity of the ground state has been carried out using magnesium and beryllium ions (\cite{03Barrett}), using beryllium and aluminium ions (\cite{05Schmidt}), with magnesium and aluminium ions (\cite{10Chou}) and using two isotopes of calcium (\cite{08Home}). Sympathetic Doppler cooling had previously been performed using cadmium ions  (\cite{02Blinov}). The use of different species of ions has the advantage that the spectral selection of one species by the laser cooling light is almost perfect, with the probability of exciting spontaneous emission from the beryllium ion during a single cycle of magnesium sideband cooling estimated at the level of 4 parts in 10$^{11}$ (\cite{03Barrett}), and the phase shift of a hyperfine superposition due to the AC Stark effect of the magnesium cooling light estimated to be $3\times 10^{-10}$~radians. This is ideal for quantum information processing where scattering would result in an error, and is thus required to be small.

The use of two isotopes offers the advantage that masses of the two ions are similar, avoiding the problems of poorly cooled modes described in section \ref{sec:cooling}. In addition similar laser systems can be used to control both ion species (which reduces cost). However the spectral addressing is not as good as for two species, because the transitions in the two different ions are separated by the isotope shift, which is of the order of a few GHz for most of the commonly trapped ion species. The limits to spectral addressing using different isotopes were examined by \cite{08Home} with \Ca{43} and \Ca{40} ions. The coherence of a qubit stored in the $\ket{^{2}S_{1/2}, F = 3, M_F = 0} \leftrightarrow \ket{^{2}S_{1/2}, F = 4, M_F = 0}$ ``clock'' states of the \Ca{43} ion was monitored as the cooling progressed. Qubit coherence was experimentally observed to be reduced by 3.3\% per cooling cycle, dominated by off-resonant photon scattering from the Raman beams used to address the motional sideband transitions of the \Ca{40} transition. By detuning the Raman laser further from resonance, this source of error would be strongly suppressed (\cite{06Ozeri}), but increased laser light intensity would be required in order to cool at the same rate.  The resonant light required to repump \Ca{40} also causes a loss of coherence in \Ca{43}, leading to a loss of fidelity of superpositions stored in the clock-state qubit with a lower limit of 1 part in 10$^4$ per cooling cycle - this is a fundamental limit due to the finite isotope shift of $\sim1$~GHz. It should be noted that in addition to photon scattering, off-resonant driving of \Ca{43} transitions by the cooling light can lead to significant phase shifts of the qubit due to the AC Stark effect, which could provide an additional source of decoherence.

\begin{figure}
\centering
\includegraphics[width=0.8\textwidth]{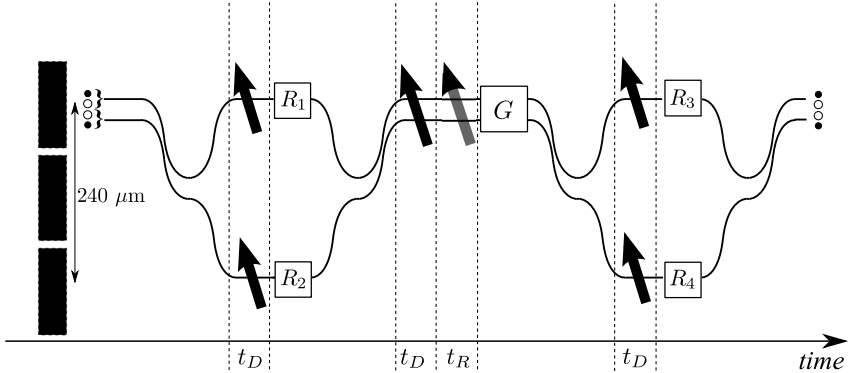}
\caption{Sequence of operations for two ``qubit$+$coolant'' packages used in the experiments reported in \cite{09Home,10Hanneke} and \cite{12Gaebler}. The sequence starts with the two magnesium-beryllium ion pairs trapped in the same zone of the trap array. These are then separated in order to perform individually addressed single qubit gates $R_1$ and $R_2$ on the qubit ions. The ions are subsequently recombined in order to perform a two-qubit logic gate $G$ between the qubits. The separation process is then repeated to enable individually addressed operations $R_3$ and $R_4$, after which all ions are brought back into the same trap in order to regain the initial ion configuration. The black arrows indicate times at which Doppler cooling was performed, and the grey arrow indicates the use of Raman sideband cooling for cooling axial modes of motion close to the ground state (enabling high-fidelity logic operations). In \cite{10Hanneke}, $t_D = 300~\mu$s and $t_R \simeq ~3$ms, and the separation and recombination times were approximately $500\mu$s.} \label{fig:Home09Sequence}
\end{figure}

Sympathetic cooling was used to demonstrate the compatibility of multi-qubit gates with transport of quantum information by \cite{09Home}, and applied to quantum information protocols including arbitrary unitary transformations on two qubits (\cite{10Hanneke}) and multi-qubit randomized benchmarking of quantum gates (\cite{12Gaebler}). In these experiments two beryllium and two magnesium ions were used, with each beryllium ``qubit'' ion accompanied at all times by an ancillary magnesium ``coolant'' ion. The beryllium ions were used to store qubits, which were manipulated using stimulated Raman transitions (\cite{98Wineland2}). An example sequence of gate and transport operations which was used by \cite{09Home} is shown in figure \ref{fig:Home09Sequence}. The magnesium ions were used for Doppler cooling of all motional modes subsequent to each transport operation, providing some level of cooling which could help in the event of severe lack of control, such as a collision with a background gas particle. Prior to each two-qubit gate, the axial motional modes were cooled close to the ground state of motion by Raman sideband cooling of the magnesium ions. The lasers used for the two-qubit gate couple only to the axial modes, so these have non-zero Lamb-Dicke parameters which could impact on the gate fidelity according to equation \ref{eq:gateerror}.  Sympathetic cooling was used to completely mitigate the effects of imperfect transport and heating. The experiments investigated a repeated sequence of the form shown in figure \ref{fig:Home09Sequence} with the single qubit operations $R_{1..4}$ set to perform the unitary  rotation $R(\pi/2, 0) \equiv \exp\left[i \pi \left( \ket{\uparrow}\bra{\downarrow} + \ket{\downarrow}\bra{\uparrow}\right)/4\right]$ of the qubits about the $x$ axis of the Bloch sphere. The two-qubit operation $G$ implements a two-qubit geometric phase gate of the type first demonstrated by \cite{03Leibfried}. In the ideal case, the sequence performs a two-qubit unitary operation, which we will here denote by  $\hat{U}$. When repeated twice this ideally results in the operation $\hat{U}^2$. In order to characterize repeatability, the researchers performed quantum process tomography (\cite{06Riebe}) on the experimental implementation of $\hat{U}$ and $\hat{U}^2$. The process deduced from the characterization of the experimental $\hat{U}$ was then mathematically repeated, and compared to the results of the experimental characterization of $\hat{U}^2$. The results were found to be consistent with each repetition of $\hat{U}$ performing exactly the same operation. This would not have been possible without the use of sympathetic cooling, since multi-qubit gate errors would increase due to the increasing lack of control due to the gain in the ions' motional energy.

\section{Quantum state engineering}\label{sec:quantumstateengineering}

\subsection{Motional state engineering}

The techniques described above for the manipulation of mixed-species ion chains have also enabled new types of quantum state engineering. \cite{09Jost} used sympathetic cooling as part of a protocol to produce a first demonstration of entanglement between remote mechanical oscillators (each oscillator being the out of phase normal mode of oscillation of a beryllium-magnesium ion pair). The protocol started with all ions in the same trap, configured with the heavier magnesium ions centered between the two beryllium ions using the re-ordering method described in section \ref{sec:reordering}. This chain was cooled to the ground state of motion of all axial modes, and a standard multi-qubit quantum logic gate was performed in order to create the entangled state $(\ket{\uparrow \downarrow} + \ket{\downarrow \uparrow})/\sqrt{2}$ of the internal states of the two beryllium ions (in this case $\ket{\uparrow} \equiv \ket{^{2}S_{1/2}, F = 2, M_F = 2}$ and $\ket{\downarrow} \equiv \ket{^{2}S_{1/2}, F = 2, M_F = 1}$). The ion crystal was then separated into two pairs, each consisting of one beryllium and one magnesium ion which were transported to two locations 240~$\mu$m apart. These locations will be labeled as $A$ and $B$ in what follows. After separation, the axial motional modes of both crystals were in an unknown mixed excited state due to imperfect control of transport. In order to create pure-state motional entanglement from the remote spin entanglement, the motional states were first sympathetically cooled into the ground state, using Doppler cooling followed by sideband cooling on both the first and second motional sidebands of the magnesium ions. The sympathetic cooling ensured that the internal state entanglement of the beryllium ions was preserved. Using a notation in which the Fock states of motion for the two normal modes of interest (in each case the out-of-phase mode of axial motion) in trap locations $A$ and $B$ are labeled $\ket{n}_A$ and $\ket{n}_B$ respectively, the combined state of the two beryllium ions and the motion is then ideally
\be
\ket{\psi_S} = \frac{1}{\sqrt{2}}\left(\ket{\downarrow}_{\rm Be1}\ket{\uparrow}_{\rm Be2} + \ket{\uparrow}_{\rm Be1} \ket{\downarrow}_{\rm Be2}\right) \ket{0}_{\rm A} \ket{0}_{\rm B} \ .
\ee
In the experiment the cooling achieved a probability of ground state occupation of $P_A(n = 0) = 0.94$ and $P_B(n = 0) = 0.98$ for the out-of-phase mode of motion of each pair of ions. Subsequent to cooling, motional sideband pulses were applied in each of the separate traps to transfer the internal state entanglement into the motional state. Though the same sideband was driven in both traps, prior to application of the sideband pulse in trap B, an inversion of the spin state was performed using the carrier transition. In the ideal case, this procedure produces the entangled motional state
\be
\ket{\psi_M} = \frac{1}{\sqrt{2}}\left(\ket{1}_{\rm A} \ket{1}_{\rm B} + \ket{0}_{\rm A} \ket{0}_{\rm B}\right) \ket{\downarrow}_{\rm Be1}\ket{\downarrow}_{\rm Be2} \ \ .
\ee
The quality with which the state was produced in the experiment was diagnosed by transferring the entanglement back into the internal state degree of freedom of the ions and evaluating the entanglement of the resulting state using standard methods (\cite{00Sackett}). This produced a lower limit of the fidelity for the motional state of 0.57(2) (\cite{09Jost}). The difference between this value and $1$ is due to a variety of factors, including imperfections in the ground state cooling, the internal state entanglement procedure and the spin-motion transfer pulses, and decoherence of the motional states before they were transferred back into the internal states.

In this protocol the sympathetic cooling was used to initialize a pure state of motion (the ground state) which was suitable for use in the subsequent quantum state engineering. This starting point could also be used to create other entangled states of remote oscillators using a similar procedure. One option would be to start from $\ket{\psi_S}$ and subsequently apply state-dependent forces to create entanglement between superpositions of coherent states at each location. These methods are well known for single ions, having been used to realize entanglement between spin and motion of single ions in a number of trapped-ion groups (see, for example \cite{96Monroe, 05Haljan, 07McDonnell}). In the case of ions in two separate traps, the entangled state produced would not involve purely motional entanglement, due to residual entanglement with the internal states.

\subsection{Entanglement of internal states}

Mixed-species ion chains have also been used to engineer entangled states of the internal degrees of freedom, making use of the difference in spectra of the two ions to address each species independently. Such methods were used by \cite{09Hume} to engineer entangled Dicke states of the internal states of two $^{25}$Mg$^+$ ions making use of a single co-trapped $^{27}$Al$^+$ ion. The $^{27}$Al$^+$ ion was stored on the end of the chain, resulting in one mode in which the ratio of the amplitudes of motion of the two $^{25}$Mg$^+$ ions is close to 1 (equal amplitudes are important for the protocol).  The internal electronic state of each ion was first initialized to one energy eigenstate by optical pumping, and the axial modes of motion initialized close to the ground state by sideband cooling the magnesium ions. Using a notation in which the initialized energy level of ion species $X$ is denoted as $\ket{\downarrow}_X$ and the second level involved in the protocol is $\ket{\uparrow}_X$, this state can be written as
\be
\ket{\psi_{\rm init}} = \ket{\downarrow}_{\rm Mg1}\ket{\downarrow}_{\rm Mg2} \ket{\downarrow}_{\rm Al} \ket{0}_{\rm mot} \ ,
\ee
where for $^{25}$Mg$^+$, $\ket{\downarrow}_{\rm Mg} \equiv \ket{^{2}S_{1/2}, F = 3, M_F = -3}$ and $\ket{\uparrow}_{\rm Mg} \equiv \ket{^{2}S_{1/2}, 2,-2}$, and for $^{27}$Al$^+$, $\ket{\downarrow}_{\rm Al} \equiv \ket{^{1}S_{0},5/2,  -5/2}$ and $\ket{\uparrow}_{\rm Al} \equiv \ket{^{3}P_{1},7/2, -7/2}$.

Starting from this initial state, the protocol involved introducing one quantum to the motion by coherently driving  the motion-adding sideband transition of the aluminium ion. In an ideal realization, the state of the three ions and the motional mode of motion is then
\be
\ket{\psi_{\rm 1}} = \ket{\downarrow}_{\rm Mg1}\ket{\downarrow}_{\rm Mg2} \ket{\uparrow}_{\rm Al} \ket{1}_{\rm mot} \ .
\ee
The final step of the protocol is then to directly drive the motion subtracting sideband of the quadrupole transition of \emph{both} magnesium ions simultaneously with equal strength. Since only one quantum can be removed from the motion, only one of the magnesium ions can change state, but there is no way of distinguishing which should change. In the ideal case the ions then evolve into a superposition of two orthogonal states where each magnesium ion changed state, resulting in
\be
\ket{\psi_{\rm 1}} = \frac{1}{\sqrt{2}}\left(\ket{\downarrow}_{\rm Mg1}\ket{\uparrow}_{\rm Mg2} + \ket{\uparrow}_{\rm Mg1}\ket{\downarrow}_{\rm Mg2}\right) \ket{\uparrow}_{\rm Al} \ket{0}_{\rm mot}
\ee
which is an entangled state of the magnesium ions. This method generalizes to create Dicke states of larger numbers of ions, so long as motional sideband transitions in all of the ions can be driven at the same rate. For mixed-species ion crystals this condition is only approximated for strings of more than three ions. Numerical studies performed by \cite{09Hume} indicate that Dicke states in which only one ion is in $\ket{\uparrow}$ and all others are $\ket{\downarrow}$ fidelities greater than 0.99 could be expected with up to 6 magnesium ions, and fidelities greater than 0.95 for states with multiple $\ket{\uparrow}$ ions in up to 6 magnesium ions. In the experiments of \cite{09Hume}, two magnesium ions were entangled with a fidelity of 77\%, primarily limited by intensity fluctuations of the control laser fields, and imperfect ground state cooling.

The protocol used by Hume et al. demonstrates the use of a second ion species for individual addressing of ions  in a chain, allowing the motional state to be engineered as part of the protocol without affecting the target species. Individual addressing has been performed using ions of the same species by tight focusing of a laser beam such that it primarily illuminates one ion in the chain. This method has been implemented with great success in quantum information work by the Innsbruck group (see for example \cite{05Haffner, 09Monz, 11Schindler}), but becomes more challenging at high trap frequencies where the spatial separation of the ions is reduced. Individual addressing using two species presents an alternative method for achieving this which holds different challenges (for instance the need for multiple laser systems, atomic sources etc.). This may be desirable in settings where tight focussing is difficult to achieve. The individual addressing provided by a second ion species has the advantage that it is simple to add resonant lasers, allowing photons to be scattered. This provides the possibility to add a combination of dissipation and coherent control of the motional modes of an ion string as an ingredient for state preparation, providing further possibilities beyond those which have been demonstrated thus far.

\section{Molecular cooling and control} \label{sec:molecules}
The quantum-logic techniques which have been demonstrated for control and readout of atomic ions which do not have accessible laser cooling transitions are also directly applicable to molecular systems. The additional degrees of freedom available in molecules hold rich promise for investigations of fundamental physics, quantum control and frequency standards. In order to perform precision spectroscopy and control of these systems it is desirable to cool all degrees of freedom to the quantum ground state. However achieving this is highly challenging, due to the relative complexity of the energy level structure of even the simplest of molecules.

Sympathetic cooling of molecular translational motion using atomic species has been reported by a number of groups, allowing Doppler cooled Coulomb crystals with large numbers of ions to be formed involving diatomic molecules such as MgH$^+$ (\cite{00Moelhave}), HD$^+$ (\cite{05Blythe}) and N$_{2}^+$ (\cite{10Tong}). One early example of molecular sympathetic cooling was the mass spectrometry performed by \cite{96Baba} which made use of resonant driven motion of the molecular species, resulting in heating of a co-trapped laser cooled atomic species through the Coulomb interaction which could be detected through reduced fluorescence (these experiments were performed in a higher temperature ``liquid'' regime, in which the ions do not crystallize to fixed equilibrium positions). Sympathetically cooled ions have been used in spectroscopy of some molecular species, but achieving high precision is very challenging due to the lack of efficient state-detection methods for interrogating the state of the molecular ion in situ. Thus the possibility of performing quantum logic spectroscopy and similar techniques for control based on these methods may open up a range of possibilities for extensions of trapped-ion quantum control into the molecular realm.

In order to take advantage of molecules for quantum state control and information processing, it is highly desirable to be able to prepare  the molecule in a single quantum state. This requires methods for isolating a single rotational and vibrational level in the molecule, and a single hyperfine state. The vibrational levels do not pose a problem in this regard, since the the energy scale of the level separations is much larger than the thermal fluctuations at 300~K. By contrast, the rotational degrees of freedom have frequencies of $\nu_{\rm rot} \sim 10^{11}$~Hz, and are thermally occupied at room temperature, with thermalization timescales on the order of tens of seconds (\cite{06Vogelius2}). One possibility for cooling the rotation would be to cool the ion trap setup itself such that the environment of the molecule fulfils $T <<  h \nu_{\rm rot}/k_B$, however this requires temperatures well below 4~K at which it seems challenging to operate an ion trap. Active cooling of rotation has been demonstrated by \cite{10Staanum} and \cite{10Schneider}, following a proposal from \cite{02Vogelius} which involves optical pumping of the first and second rotational excited states into a higher vibrational level, from which there is some chance of decay into the ro-vibrational ground state (which is a dark state for the pumping laser due to frequency selection). Population transfer from higher rotational states relies on blackbody radiation, which limits the speed of the pumping, and thus the temperature of the final steady state. \cite{12Bressel} built upon the ability to perform rotational cooling, using an optical frequency comb to prepare a particular hyperfine state of HD$^{+}$ molecules by driving multiple internal state molecular transitions. Rather than trying to perform state preparation on the ions themselves, \cite{10Tong} have demonstrated rotational state preparation by state-selective photo-ionization of molecular atoms.

A Raman cooling method similar to the internal state preparation described in section \ref{sec:quantumlogicreadout} provides an alternative approach to cooling the rotational degree of freedom (\cite{06Vogelius}, \cite{06Schmidt}). The method proposed by \cite{06Vogelius} is based on driving Raman transitions between the $\ket{J = 2}_{\rm rot}\ket{\nu}_{\rm trans} \leftrightarrow \ket{J = 0}_{\rm rot}\ket{\nu'}_{\rm trans}$ states, where $\nu$ is the translational quantum number and $J$ is the rotational quantum number. If the Raman difference frequency is $2 \omega_{\rm rot} + (\delta\nu) \omega_{\rm trans}$, the transition is resonant for $\delta \nu = \nu' - \nu$. Choosing a Raman laser detuning for which $\delta\nu < 0$ the resonant transitions reduce the translational quantum number as the rotational quantum number is increased by two, and vice versa. The second component of the scheme is continuous cooling of the collective translational motion, which is performed using the atomic ion (either by Doppler or Raman sideband cooling). If the translational cooling is fast compared to the ro-translational Raman transition, the occupied translational states are most likely to be those with $\nu, \nu'$ close to zero (the ground state). For translational states for which $\nu' < -\delta \nu $, no lower translational state is available, biasing the relative rates of the Raman transitions between $J = 0$ and $J=2$ towards transitions which reduce the rotational quantum number. The result is a unidirectional pumping from $\ket{J = 2} \rightarrow \ket{J = 0}$. It should be noted that the Raman transition has a linewidth $\Gamma$ given by the damping of the translational levels due to the cooling light, and the natural linewidth. In order for the scheme to succeed, this is required to satisfy $\Gamma < \delta \nu \omega_{\rm trans}$ in order that the desired selectivity of transitions towards those that reduce the translational quantum number can be achieved.

The dissipative step in the rotational cooling is provided by the emission of the photon from the atomic ion into the vacuum during the cooling of the translational motion. The Raman cooling in this proposal is of a single rotational degree of freedom. Other rotational degrees of freedom could potentially be cooled directly using different transition frequencies, but are also coupled to the actively cooled degree of freedom by black-body radiation. This process happens on timescales on the order of 10~s, which are long compared to typical timescales for translational cooling using either Doppler or Raman sideband methods. \cite{06Vogelius} performed simulations with the atomic ion being Doppler cooled, which indicate that it may be possible to cool the rotational motion to 80~$\%$ occupancy of the $J = 1$ and $J = 0$ levels in around 40~s, with $>60\%$ occupancy of the $J = 0$ state. The simulations assumed a very high translational frequency of 50~MHz, which is due to the requirement that the scheme requires $\delta \nu \omega_{\rm trans} > \Gamma$. The trap frequency could likely be reduced if the translational cooling was performed using cooling on a narrow linewidth transition, which could also allow the ro-vibrational-translational ground state to be reached with higher probability.

A second paper by the same authors (\cite{06Vogelius2}) proposed probabilistic state preparation of molecules using quantum logic spectroscopy, along the lines used for internal state preparation by \cite{05Schmidt}. The translational normal modes are cooled to the ground state, and the molecule is subsequently probed using a process similar to stimulated Raman adiabatic passage (STIRAP) which adds a quantum to the motion if the ion is in a rotational state which is close to resonance with the STIRAP lasers. This results in excitation of the translational motion which can be detected on the atomic ion. \cite{07Koelemeij} proposed a similar scheme involving a coherent drive on the molecular ion produced by state-dependent forces. The latter is similar to one of the methods demonstrated by \cite{11Hume} on atomic ions and described in section \ref{sec:QLogicCoherent} above. State-dependent force methods resulting in geometric phases which are detected on the atomic species have also been proposed by \cite{12MurPetit}, including quantum gates between the atomic and molecular ion species.

A challenging point in proposals involving molecular ions is that driving Raman transitions between states which are spaced by $\sim 10^{11}$~Hz using continuous-wave lasers requires phase-locked frequency components spaced by this frequency. In performing initial spectroscopy of molecular ions (which is necessary because for many molecules the structures are not currently known), it is desirable that many different possible rotational Raman transitions of the molecule could be probed, requiring a large range of frequencies. Optical frequency combs (\cite{03Cundiff}) provide exactly such a set of phase coherent frequency components, and were recently used to drive Raman transitions between hyperfine states in atomic ions by \cite{10Hayes}. The comb produces frequency components at $\omega_{\rm offset} + n \omega_{\rm rep}$. By splitting the comb into two paths, and shifting the frequency of one by $\omega_{\rm shift}$ relative to the other, it is possible to produce a set of Raman transitions with frequencies
\be
\omega_{\rm Raman} = \Delta n \omega_{\rm rep} + \omega_{\rm shift}
\ee
where $\Delta n$ ranges from $0$ up to the ratio of the bandwidth of the comb to the repetition frequency. It should be noted that contributions from all pairs of comb teeth with frequency difference $\omega_{\rm Raman}$ add, so that the transition rates make use of a significant fraction of the power of the comb laser (\cite{10Hayes}). The range of different $\Delta n$ obtained from a single laser source provides the possibility to access a large range of transitions (the exact resonance condition must be met by tuning $\omega_{\rm shift}$ using acousto-optic-modulators or similar devices). This was used in the work of \cite{12Bressel} on HD$^+$ spectroscopy. The use of a frequency comb forms a key ingredient in two recent proposals by \cite{12Leibfried} and \cite{12Ding} for performing a systematic search of the transitions in a molecular ion, assuming little or no prior knowledge of its internal state. In both papers, quantum logic readout using translationally ground-state cooled ions is combined with the ability to probe the molecular ion with a large range of Raman frequency components derived from a frequency comb.

\section{Outlook} \label{sec:conclusion}

The quantum manipulation of mixed-species chains of trapped ions is a relatively recent development which has required a number of new considerations for ion trap physics. A much better understanding of the trapping potential is required, due to the difference in masses of the different species. New techniques have been demonstrated to control these systems, resulting in a number of new opportunities for the control and preparation of quantum states. It is expected that additional techniques will emerge in the near future. For instance, the ability to manipulate the relative components of dissipation, measurement and coherent control available in these systems seems likely to make them particularly amenable to dissipative state preparation protocols and to quantum simulation of open quantum systems (\cite{11Barreiro}). Quantum logic spectroscopy has already provided the leading frequency standard to date, and is likely to open up quantum control of trapped ions to a wide range of molecular species, providing new avenues for research.

\section{Acknowledgements}
I would like to thank David Hume for useful correspondence regarding the aluminium clock experiments, and Piet Schmidt, Stefan Willitsch, Joseba Alonso and John Jost for helpful comments on the manuscript.





\bibliography{../../../Publications/References/myrefs}

\begin{thebibliography}{89}
\expandafter\ifx\csname natexlab\endcsname\relax\def\natexlab#1{#1}\fi
\expandafter\ifx\csname url\endcsname\relax
  \def\url#1{\texttt{#1}}\fi
\expandafter\ifx\csname urlprefix\endcsname\relax\def\urlprefix{URL }\fi

\bibitem[{Allcock et~al.(2011)Allcock, Guidoni, Harty, Ballance, Blain, Steane,
  and Lucas}]{11Allcock}
Allcock, D., Guidoni, L., Harty, T.~P., Ballance, C., Blain, M., Steane, A.,
  Lucas, D., 2011. Reduction of heating rate in a microfabricated ion trap by
  pulsed-laser cleaning. New J. Phys. 13, 123023.

\bibitem[{Amini et~al.(2010)Amini, Uys, Wesenberg, Seidelin, Britton,
  Bollinger, Leibfried, Ospelkaus, VanDevender, and Wineland}]{10Amini}
Amini, J.~M., Uys, H., Wesenberg, J.~H., Seidelin, S., Britton, J., Bollinger,
  J.~J., Leibfried, D., Ospelkaus, C., VanDevender, A.~P., Wineland, D.~J.,
  2010. Toward scalable ion traps for quantum information processing. New. J.
  Phys. 12, 033031.

\bibitem[{Baba and Waki(1996)}]{96Baba}
Baba, T., Waki, I., 1996. Cooling and mass-analysis of molecules using
  laser-cooled atoms. JAPANESE JOURNAL OF APPLIED PHYSICS PART 2 LETTERS 35,
  1134--1137.

\bibitem[{Barreiro et~al.(2011)Barreiro, M{\"u}ller, Schindler, Nigg, Monz,
  Chwalla, Hennrich, Roos, Zoller, and Blatt}]{11Barreiro}
Barreiro, J.~T., M{\"u}ller, M., Schindler, P., Nigg, D., Monz, T., Chwalla,
  M., Hennrich, M., Roos, C.~F., Zoller, P., Blatt, R., 2011. An open-system
  quantum simulator with trapped ions. Nature 470~(7335), 486--491.

\bibitem[{Barrett et~al.(2003)Barrett, DeMarco, Schaetz, Leibfried, Britton,
  Chiaverini, Itano, Jelenkovic, Jost, Langer, Rosenband, and
  Wineland}]{03Barrett}
Barrett, M.~D., DeMarco, B., Schaetz, T., Leibfried, D., Britton, J.,
  Chiaverini, J., Itano, W.~M., Jelenkovic, B., Jost, J.~D., Langer, C.,
  Rosenband, T., Wineland, D.~J., 2003. Sympathetic cooling of {Be}$^+$ and
  {Mg}$^+$ for quantum logic. Phys. Rev. A 68~(042302).

\bibitem[{Benhelm et~al.(2008)Benhelm, Kirchmair, Roos, and Blatt}]{08Benhelm}
Benhelm, J., Kirchmair, G., Roos, C.~F., Blatt, R., 2008. Towards
  fault-tolerant quantum computing with trapped ions. Nature Physics 4, 463 --
  466.

\bibitem[{Berkeland et~al.(1998{\natexlab{a}})Berkeland, Miller, Bergquist,
  Itano, and Wineland}]{98Berkeland}
Berkeland, D.~J., Miller, J.~D., Bergquist, J.~C., Itano, W.~M., Wineland,
  D.~J., 1998{\natexlab{a}}. Laser cooled mercury ion frequency standard. Phys.
  Rev. Lett. 80, 2089.

\bibitem[{Berkeland et~al.(1998{\natexlab{b}})Berkeland, Miller, Bergquist,
  Itano, and Wineland}]{98Berkeland2}
Berkeland, D.~J., Miller, J.~D., Bergquist, J.~C., Itano, W.~M., Wineland,
  D.~J., 1998{\natexlab{b}}. Minimization of ion micromotion in a {P}aul trap.
  Journal of Applied Physics 83~(10), 5025--5033.

\bibitem[{Biercuk et~al.(2010)Biercuk, Uys, Britton, VanDevender, and
  Bollinger}]{10Biercuk}
Biercuk, M., Uys, H., Britton, J., VanDevender, A., Bollinger, J.~J., 2010.
  Ultrasensitive force and displacement detection using trapped ions. Nature
  Nanotechnology 5, 646--650.

\bibitem[{Blakestad et~al.(2009)Blakestad, Ospelkaus, VanDevender, Amini,
  Britton, Leibfried, and Wineland}]{09Blakestad}
Blakestad, R.~B., Ospelkaus, C., VanDevender, A.~P., Amini, J.~M., Britton, J.,
  Leibfried, D., Wineland, D.~J., 2009. High-fidelity transport of trapped-ion
  qubits through an x-junction trap array. Phys. Rev. Lett. 102, 153002.

\bibitem[{Blatt and Wineland(2008)}]{08Blatt}
Blatt, R., Wineland, D.~J., 2008. Entangled states of trapped atomic ions.
  Nature 453, 1008--1015.

\bibitem[{Blinov et~al.(2002)Blinov, Deslauriers, Madsen, Miller, and
  Monroe}]{02Blinov}
Blinov, B., Deslauriers, L., Madsen, M.~J., Miller, R., Monroe, C., 2002.
  Sympathetic cooling of trapped {C}d isotopes. Phys. Rev. A 65, 040304.

\bibitem[{Blythe et~al.(2005)Blythe, Roth, Fr\"ohlich, Wenz, and
  Schiller}]{05Blythe}
Blythe, P., Roth, B., Fr\"ohlich, U., Wenz, H., Schiller, S., Oct 2005.
  Production of ultracold trapped molecular hydrogen ions. Phys. Rev. Lett. 95,
  183002.
\newline\urlprefix\url{http://link.aps.org/doi/10.1103/PhysRevLett.95.183002}

\bibitem[{Bressel et~al.(2012)Bressel, Borodin, Shen, Hansen, Ernsting, and
  Schiller}]{12Bressel}
Bressel, U., Borodin, A., Shen, J., Hansen, M., Ernsting, I., Schiller, S., May
  2012. Manipulation of individual hyperfine states in cold trapped molecular
  ions and application to ${\mathrm{hd}}^{+}$ frequency metrology. Phys. Rev.
  Lett. 108, 183003.
\newline\urlprefix\url{http://link.aps.org/doi/10.1103/PhysRevLett.108.183003}

\bibitem[{Chou et~al.(2010)Chou, Hume, Koelemeij, Wineland, and
  Rosenband}]{10Chou}
Chou, C.~W., Hume, D.~B., Koelemeij, J. C.~J., Wineland, D.~J., Rosenband, T.,
  2010. Frequency comparison of two high-accuracy {Al}$^+$ optical clocks.
  Phys. Rev. Lett. 104, 070802.

\bibitem[{Cundiff and Ye(2003)}]{03Cundiff}
Cundiff, S.~T., Ye, J., Mar 2003. \textit{Colloquium} : Femtosecond optical
  frequency combs. Rev. Mod. Phys. 75, 325--342.
\newline\urlprefix\url{http://link.aps.org/doi/10.1103/RevModPhys.75.325}

\bibitem[{Daniilidis et~al.(2011)Daniilidis, Narayanan, Möller, Clark, Lee,
  Leek, Wallraff, Schulz, Schmidt-Kaler, and Häffner}]{11Danilidis}
Daniilidis, N., Narayanan, S., Möller, S.~A., Clark, R., Lee, T.~E., Leek,
  P.~J., Wallraff, A., Schulz, S., Schmidt-Kaler, F., Häffner, H., 2011.
  Fabrication and heating rate study of microscopic surface electrode ion
  traps. New Journal of Physics 13~(1), 013032.
\newline\urlprefix\url{http://stacks.iop.org/1367-2630/13/i=1/a=013032}

\bibitem[{Dehmelt(1969)}]{69Dehmelt}
Dehmelt, H.~G., 1969. Radiofrequency {s}pectroscopy of stored ions {II}. Adv.
  Atom. Mol. Phys. 5, 109.

\bibitem[{Dehmelt(1990)}]{90Dehmelt}
Dehmelt, H.~G., 1990. Experiments with an isolated subatomic particle at rest.
  Rev. Mod. Phys. 62~(3), 525--530.

\bibitem[{Deslauriers et~al.(2006)Deslauriers, Olmschenk, Stick, Hensinger,
  Sterk, and Monroe}]{06Deslauriers}
Deslauriers, L., Olmschenk, S., Stick, D., Hensinger, W.~K., Sterk, J., Monroe,
  C., Sep 2006. Scaling and suppression of anomalous heating in ion traps.
  Phys. Rev. Lett. 97, 103007.
\newline\urlprefix\url{http://link.aps.org/doi/10.1103/PhysRevLett.97.103007}

\bibitem[{Ding and Matsukevich(2012)}]{12Ding}
Ding, S., Matsukevich, D.~N., 2012. Quantum logic for the control and
  manipulation of molecular ions using a frequency comb. New Journal of Physics
  14~(2), 023028.
\newline\urlprefix\url{http://stacks.iop.org/1367-2630/14/i=2/a=023028}

\bibitem[{Drullinger et~al.(1980)Drullinger, Wineland, , and
  Bergquist}]{80Drullinger}
Drullinger, R.~E., Wineland, D. .~J., , Bergquist, J.~C., 1980. High-
  resolution optical spectra of laser cooled ions. App. Phys. 22, 365--368.

\bibitem[{Gaebler et~al.(2012)Gaebler, Meier, Tan, Bowler, Lin, Hanneke, Jost,
  Home, Knill, Leibfried, and Wineland}]{12Gaebler}
Gaebler, J.~P., Meier, A.~M., Tan, T.~R., Bowler, R., Lin, Y., Hanneke, D.,
  Jost, J.~D., Home, J.~P., Knill, E., Leibfried, D., Wineland, D.~J., 2012.
  Randomized benchmarking of multiqubit gates. Phys. Rev. Lett. 108, 260503.

\bibitem[{Ghosh(1995)}]{BkGhosh}
Ghosh, P.~K., 1995. Ion Traps. Oxford University Press.

\bibitem[{Gottesman and Chuang(1999)}]{99Gottesman}
Gottesman, D., Chuang, I.~L., 1999. Quantum teleportation is a universal
  computational primitive. Nature 402, 390, quant-ph/9908010.

\bibitem[{H\"{a}ffner et~al.(2005)H\"{a}ffner, H{\"a}nsel, Roos, Benhelm,
  al~kar, Chwalla, K\"{o}rber, Rapol, Riebe, Schmidt, Becher, G\"{u}hne,
  D\"{u}r, and Blatt}]{05Haffner}
H\"{a}ffner, H., H{\"a}nsel, W., Roos, C.~F., Benhelm, J., al~kar, D.~C.,
  Chwalla, M., K\"{o}rber, T., Rapol, U.~D., Riebe, M., Schmidt, P.~O., Becher,
  C., G\"{u}hne, O., D\"{u}r, W., Blatt, R., 2005. Scalable multiparticle
  entanglement of trapped ions. Nature 438, 643.

\bibitem[{Haljan et~al.(2005)Haljan, Lee, Brickman, Acton, Deslauriers, and
  Monroe}]{05Haljan}
Haljan, P.~C., Lee, P.~J., Brickman, K.-A., Acton, M., Deslauriers, L., Monroe,
  C., 2005. Entanglement of trapped-ion clock states. Phys. Rev. A 72, 062316.

\bibitem[{Hanneke et~al.(2009)Hanneke, Home, Jost, Amini, Leibfried, and
  Wineland}]{10Hanneke}
Hanneke, D., Home, J.~P., Jost, J.~D., Amini, J.~M., Leibfried, D., Wineland,
  D.~J., 2009. Realization of a programmable two-qubit quantum processor.
  Nature Physics 6, 13--16.

\bibitem[{Hayes et~al.(2010)Hayes, Matsukevich, Maunz, Hucul, Quraishi,
  Olmschenk, Campbell, Mizrahi, Senko, and Monroe}]{10Hayes}
Hayes, D., Matsukevich, D.~N., Maunz, P., Hucul, D., Quraishi, Q., Olmschenk,
  S., Campbell, W., Mizrahi, J., Senko, C., Monroe, C., Apr 2010. Entanglement
  of atomic qubits using an optical frequency comb. Phys. Rev. Lett. 104,
  140501.
\newline\urlprefix\url{http://link.aps.org/doi/10.1103/PhysRevLett.104.140501}

\bibitem[{Hensinger et~al.(2006)Hensinger, Olmschenk, Stick, Hucul, Yeo, Acton,
  Deslauriers, Monroe, and Rabchuk}]{06Hensinger}
Hensinger, W.~K., Olmschenk, S., Stick, D., Hucul, D., Yeo, M., Acton, M.,
  Deslauriers, L., Monroe, C., Rabchuk, J., 2006. T-junction ion trap array for
  two-dimensional ion shuttling, storage, and manipulation. Applied Physics
  Letters 88~(3), 034101.
\newline\urlprefix\url{http://link.aip.org/link/?APL/88/034101/1}

\bibitem[{Hite et~al.(2012)Hite, Colombe, Wilson, Brown, Warring, J\"ordens,
  Jost, McKay, Pappas, Leibfried, and Wineland}]{12Hite}
Hite, D.~A., Colombe, Y., Wilson, A.~C., Brown, K.~R., Warring, U., J\"ordens,
  R., Jost, J.~D., McKay, K.~S., Pappas, D.~P., Leibfried, D., Wineland, D.~J.,
  Sep 2012. 100-fold reduction of electric-field noise in an ion trap cleaned
  with \textit{In~Situ} argon-ion-beam bombardment. Phys. Rev. Lett. 109,
  103001.
\newline\urlprefix\url{http://link.aps.org/doi/10.1103/PhysRevLett.109.103001}

\bibitem[{Home et~al.(2009{\natexlab{a}})Home, Hanneke, Jost, Amini, Leibfried,
  and Wineland}]{09Home}
Home, J.~P., Hanneke, D., Jost, J.~D., Amini, J.~M., Leibfried, D., Wineland,
  D.~J., 2009{\natexlab{a}}. Complete methods set for scalable ion trap quantum
  information processing. Science 325, 1227.

\bibitem[{Home et~al.(2011)Home, Hanneke, Jost, Leibfried, and
  Wineland}]{11Home}
Home, J.~P., Hanneke, D., Jost, J.~D., Leibfried, D.~I., Wineland, D.~J., 2011.
  Normal modes of trapped ions in the presence of anharmonic trapping
  potentials. New. J. Phys. 13, 073026.

\bibitem[{Home et~al.(2006)Home, McDonnell, Lucas, Imreh, Keitch, Szwer,
  Thomas, Stacey, and Steane}]{06Home}
Home, J.~P., McDonnell, M.~J., Lucas, D.~M., Imreh, G., Keitch, B.~C., Szwer,
  D.~J., Thomas, N.~R., Stacey, D.~N., Steane, A.~M., 2006. Entanglement and
  tomography of ion spin qubits. New J. Phys. 8, 188.

\bibitem[{Home et~al.(2009{\natexlab{b}})Home, McDonnell, Szwer, Keitch, Lucas,
  Stacey, and Steane}]{08Home}
Home, J.~P., McDonnell, M.~J., Szwer, D.~J., Keitch, B.~C., Lucas, D.~M.,
  Stacey, D.~N., Steane, A.~M., 2009{\natexlab{b}}. Memory coherence of a
  sympathetically cooled trapped-ion qubit. Phys. Rev. A 79, 050305(R).

\bibitem[{Hume(2010)}]{10HumeThesis}
Hume, D., 2010. Two species ion chains for quantum logic spectroscopy and
  entanglement generation. Ph.D. thesis, University of Colorado.

\bibitem[{Hume et~al.(2011)Hume, Chou, Leibrandt, Thorpe, Wineland, and
  Rosenband}]{11Hume}
Hume, D.~B., Chou, C.~W., Leibrandt, D.~R., Thorpe, M.~J., Wineland, D.~J.,
  Rosenband, T., 2011. Trapped-ion state detection through coherent motion.
  Phys. Rev. Lett. 107, 243902.

\bibitem[{Hume et~al.(2009)Hume, Chou, Rosenband, and Wineland}]{09Hume}
Hume, D.~B., Chou, C.~W., Rosenband, T., Wineland, D.~J., 2009. Preparation of
  {Dicke} states in an ion chain. Phys. Rev. A 80, 052302.

\bibitem[{Hume et~al.(2007)Hume, Rosenband, and Wineland}]{07Hume}
Hume, D.~B., Rosenband, T., Wineland, D.~J., 2007. High-fidelity adaptive qubit
  detection through repetitive quantum nondemolition measurements. Phys. Rev.
  Lett. 99, 120502.

\bibitem[{James(1998)}]{98James1}
James, D. F.~V., 1998. Quantum dynamics of cold trapped ions with application
  to quantum computation. Applied Physics B 66, 181--190.

\bibitem[{Jost et~al.(2009)Jost, Home, Amini, Hanneke, Ozeri, Langer,
  Bollinger, Leibfried, and Wineland}]{09Jost}
Jost, J.~D., Home, J.~P., Amini, J.~M., Hanneke, D., Ozeri, R., Langer, C.,
  Bollinger, J.~J., Leibfried, D., Wineland, D.~J., 2009. Entangled mechanical
  oscillators. Nature 459, 683--685.

\bibitem[{Kielpinski et~al.(2002)Kielpinski, C.Monroe, and
  Wineland}]{02Kielpinski}
Kielpinski, D., C.Monroe, Wineland, D., 2002. Architecture for a large-scale
  ion-trap quantum computer. Nature 417, 709--711.

\bibitem[{Kielpinski et~al.(2000)Kielpinski, King, Myatt, Sackett, Turchette,
  Itano, Monroe, and Wineland}]{00Kielpinski}
Kielpinski, D., King, B.~E., Myatt, C.~J., Sackett, C.~A., Turchette, Q.~A.,
  Itano, W.~M., Monroe, C., Wineland, D.~J., 2000. Sympathetic cooling of
  trapped ions for quantum logic. Phys. Rev. A 61~(032310).

\bibitem[{Kirchmair et~al.(2009)Kirchmair, Benhelm, Zähringer, Gerritsma,
  Roos, and Blatt}]{09Kirchmair}
Kirchmair, G., Benhelm, J., Zähringer, F., Gerritsma, R., Roos, C.~F., Blatt,
  R., 2009. Deterministic entanglement of ions in thermal states of motion. New
  Journal of Physics 11~(2), 023002.
\newline\urlprefix\url{http://stacks.iop.org/1367-2630/11/i=2/a=023002}

\bibitem[{Knill(2005)}]{05Knill}
Knill, E., 2005. Quantum computation with realistically noisy devices. Nature
  434, 39--44.

\bibitem[{Koelemeij et~al.(2007)Koelemeij, Roth, and Schiller}]{07Koelemeij}
Koelemeij, J. C.~J., Roth, B., Schiller, S., Aug 2007. Blackbody thermometry
  with cold molecular ions and application to ion-based frequency standards.
  Phys. Rev. A 76, 023413.
\newline\urlprefix\url{http://link.aps.org/doi/10.1103/PhysRevA.76.023413}

\bibitem[{Labaziewicz et~al.(2008)Labaziewicz, Ge, Antohi, Leibrandt, Brown,
  and Chuang}]{08Labaziewicz}
Labaziewicz, J., Ge, Y., Antohi, P., Leibrandt, D., Brown, K.~R., Chuang,
  I.~L., 2008. Suppression of heating rates in cryogenic surface-electrode ion
  traps. Phys. Rev. Lett. 100, 013001.

\bibitem[{Landau and Lifshitz(1976)}]{LLmechanics}
Landau, L.~D., Lifshitz, E.~M., 1976. Mechanics, 3rd Edition.
  Butterworth-Heinenann.

\bibitem[{Larson et~al.(1986)Larson, Bergquist, Bollinger, Itano, and
  Wineland}]{86Larson}
Larson, D.~J., Bergquist, J.~C., Bollinger, J.~J., Itano, W.~M., Wineland,
  D.~J., 1986. Sympathetic cooling of trapped ions in a laser cooled
  two--species nonneutral ion plasma. Phys. Rev. Lett. 57, 70.

\bibitem[{Leibfried(2012)}]{12Leibfried}
Leibfried, D., 2012. Quantum state preparation and control of single molecular
  ions. New Journal of Physics 14~(2), 023029.
\newline\urlprefix\url{http://stacks.iop.org/1367-2630/14/i=2/a=023029}

\bibitem[{Leibfried et~al.(2003{\natexlab{a}})Leibfried, Blatt, Monroe, and
  Wineland}]{03Leibfried2}
Leibfried, D., Blatt, R., Monroe, C., Wineland, D., 2003{\natexlab{a}}. Quantum
  dynamics of single trapped ions. Rev. Mod. Phys. 75, 281--324.

\bibitem[{Leibfried et~al.(2003{\natexlab{b}})Leibfried, DeMarco, Meyer, Lucas,
  Barrett, Britton, Itano, Jelenkovic, Langer, Rosenband, and
  Wineland}]{03Leibfried}
Leibfried, D., DeMarco, B., Meyer, V., Lucas, D., Barrett, M., Britton, J.,
  Itano, W.~M., Jelenkovic, B., Langer, C., Rosenband, T., Wineland, D.~J.,
  2003{\natexlab{b}}. Experimental demonstration of a robust and high-fidelity
  geometric two ion-qubit phase gate. Nature 422, 412--415.

\bibitem[{McDonnell et~al.(2007)McDonnell, Home, Lucas, Imreh, Keitch, Szwer,
  Thomas, Webster, Stacey, , and Steane}]{07McDonnell}
McDonnell, M.~J., Home, J.~P., Lucas, D.~M., Imreh, G., Keitch, B.~C., Szwer,
  D.~J., Thomas, N.~R., Webster, S.~C., Stacey, D.~N., , Steane, A.~M., 2007.
  Long-lived mesoscopic entanglement outside the {Lamb}-{Dicke} regime. Phys.
  Rev. Lett. 98, 063603.

\bibitem[{Metcalf and van~der Straten(1999)}]{BkMetcalf}
Metcalf, H.~J., van~der Straten, P., 1999. Laser cooling and trapping.
  Springer-Verlag, New York.

\bibitem[{Moehring et~al.(2007)Moehring, Maunz, Olmschenk, Younge, Matsukevich,
  Duan, and Monroe}]{07Moehring}
Moehring, D.~L., Maunz, P., Olmschenk, S., Younge, K.~C., Matsukevich, D.~N.,
  Duan, L.-M., Monroe, C., 2007. Entanglement of single-atom quantum bits at a
  distance. Nature 449, 68.

\bibitem[{M\o{}lhave and Drewsen(2000)}]{00Moelhave}
M\o{}lhave, K., Drewsen, M., Jun 2000. Formation of translationally cold
  ${\mathrm{mgh}}^{+}$ and ${\mathrm{mgd}}^{+}$ molecules in an ion trap. Phys.
  Rev. A 62, 011401.
\newline\urlprefix\url{http://link.aps.org/doi/10.1103/PhysRevA.62.011401}

\bibitem[{Monroe et~al.(1996)Monroe, Meekhof, King, and Wineland}]{96Monroe}
Monroe, C., Meekhof, D.~M., King, B.~E., Wineland, D.~J., 1996. A
  {Schr{\"o}dinger} cat superposition state of an atom. Science 272, 1131.

\bibitem[{Monz et~al.(2009)Monz, Kim, Villar, Schindler, Chwalla, Riebe, Roos,
  H{\"a}ffner, H{\"a}nsel, Hennrich, and Blatt}]{09Monz}
Monz, T., Kim, K., Villar, A.~S., Schindler, P., Chwalla, M., Riebe, M., Roos,
  C.~F., H{\"a}ffner, H., H{\"a}nsel, W., Hennrich, M., Blatt, R., 2009.
  Realization of universal ion-trap quantum computation with decoherence-free
  qubits. Phys. Rev. Lett. 103, 200503.

\bibitem[{Morigi and Walther(2001)}]{01Morigi}
Morigi, G., Walther, H., 2001. Two species {Coulomb} chains for quantum
  information. Eur. Phys. Jour. D 13~(261-269).

\bibitem[{Mur-Petit et~al.(2012)Mur-Petit, Garc\'{\i}a-Ripoll,
  P\'{e}rez-R\'{\i}os, Campos-Mart\'{\i}nez, Hern\'{a}ndez, and
  Willitsch}]{12MurPetit}
Mur-Petit, J., Garc\'{\i}a-Ripoll, J.-J., P\'{e}rez-R\'{\i}os, J.,
  Campos-Mart\'{\i}nez, J., Hern\'{a}ndez, M.~I., Willitsch, S., Feb 2012.
  Temperature-independent quantum logic for molecular spectroscopy. Phys. Rev.
  A 85, 022308.

\bibitem[{Ospelkaus et~al.(2011)Ospelkaus, Warring, Colombe, Brown, Amini,
  Leibfried, and Wineland}]{11Ospelkaus}
Ospelkaus, C., Warring, U., Colombe, Y., Brown, K.~R., Amini, J.~M., Leibfried,
  D., Wineland, D.~J., 2011. Microwave quantum logic gates for trapped ions.
  Nature 476, 181--184.

\bibitem[{Ozeri et~al.(2007)Ozeri, Itano, Blakestad, Britton, Chiaverini, Jost,
  Langer, Leibfried, Reichle, Seidelin, Wesenburg, and Wineland}]{06Ozeri}
Ozeri, R., Itano, W.~M., Blakestad, R.~B., Britton, J., Chiaverini, J., Jost,
  J.~D., Langer, C., Leibfried, D., Reichle, R., Seidelin, S., Wesenburg,
  J.~H., Wineland, D.~J., 2007. Errors in trapped-ion quantum gates due to
  spontaneous photon scattering. Phys. Rev. A 75, 4.

\bibitem[{Poschinger et~al.(2010)Poschinger, Walther, Singer, and
  Schmidt-Kaler}]{10Poschinger}
Poschinger, U., Walther, A., Singer, K., Schmidt-Kaler, F., Dec 2010. Observing
  the phase space trajectory of an entangled matter wave packet. Phys. Rev.
  Lett. 105, 263602.
\newline\urlprefix\url{http://link.aps.org/doi/10.1103/PhysRevLett.105.263602}

\bibitem[{Riebe et~al.(2006)Riebe, Kim, Schindler, Monz, Schmidt, K{\"o}rber,
  H{\"a}nsel, H{\"a}ffner, Roos, and Blatt}]{06Riebe}
Riebe, M., Kim, K., Schindler, P., Monz, T., Schmidt, P.~O., K{\"o}rber, T.~K.,
  H{\"a}nsel, W., H{\"a}ffner, H., Roos, C.~F., Blatt, R., 2006. Process
  tomography of ion trap quantum gates. Phys. Rev. Lett. 97, 220407.

\bibitem[{Roos(2000)}]{ThRoos}
Roos, C., 2000. Controlling the quantum state of trapped ions. Ph.{D}.~thesis,
  Universit\"{a}t Innsbruck, Austria.

\bibitem[{Rosenband et~al.(2007)Rosenband, Schmidt, Hume, Itano, Fortier,
  Stalnaker, Kim, Diddams, Koelemeij, Bergquist, and Wineland}]{07Rosenband}
Rosenband, T., Schmidt, P.~O., Hume, D.~B., Itano, W.~M., Fortier, T.~M.,
  Stalnaker, J.~E., Kim, K., Diddams, S.~A., Koelemeij, J. C.~J., Bergquist,
  J.~C., Wineland, D.~J., 2007. Observation of the $^{1}${S}$_{0} \rightarrow
  {}^{3}${P}$_{0}$ clock transition in $^{27}\mathrm{Al}^{+}$. Phys. Rev. Lett.
  98, 220801.

\bibitem[{Rowe et~al.(2002)Rowe, Ben-Kish, DeMarco, Leibfried, Meyer, Beall,
  Britton, Hughes, Itano, Jelenkovi\'{c}, Langer, Rosenband, and
  Wineland}]{02Rowe}
Rowe, M.~A., Ben-Kish, A., DeMarco, B., Leibfried, D., Meyer, V., Beall, J.,
  Britton, J., Hughes, J., Itano, W.~M., Jelenkovi\'{c}, B., Langer, C.,
  Rosenband, T., Wineland, D.~J., 2002. Transport of quantum states and
  separation of ions in a dual rf ion trap. Quantum Information and Computation
  2, 257.

\bibitem[{Sackett et~al.(2000)Sackett, Kielpinski, King, Meyer, Myatt, Rowe,
  amd W.~M.~Itano, Wineland, and Monroe}]{00Sackett}
Sackett, C.~A., Kielpinski, D., King, B.~E., Meyer, C. L.~V., Myatt, C.~J.,
  Rowe, M., amd W.~M.~Itano, Q. A.~T., Wineland, D.~J., Monroe, C., 2000.
  Experimental entanglement of four particles. Nature 404, 256.

\bibitem[{Safavi-Naini et~al.(2011)Safavi-Naini, Rabl, Weck, and
  Sadeghpour}]{11Safavi}
Safavi-Naini, A., Rabl, P., Weck, P.~F., Sadeghpour, H.~R., Aug 2011.
  Microscopic model of electric-field-noise heating in ion traps. Phys. Rev. A
  84, 023412.
\newline\urlprefix\url{http://link.aps.org/doi/10.1103/PhysRevA.84.023412}

\bibitem[{Schindler et~al.(2011)Schindler, Barreiro, Monz, Nebendahl, Nigg,
  Chwalla, Hennrich, and Blatt}]{11Schindler}
Schindler, P., Barreiro, J.~T., Monz, T., Nebendahl, V., Nigg, D., Chwalla, M.,
  Hennrich, M., Blatt, R., 2011. Experimental repetitive quantum error
  corrrection. Science 332, 1059--1061.

\bibitem[{Schmidt et~al.(2006)Schmidt, Rosenband, Koelemeij, Hume, Itano,
  Bergquist, and Wineland}]{06Schmidt}
Schmidt, P.~O., Rosenband, T., Koelemeij, J. C.~J., Hume, D.~B., Itano, W.~M.,
  Bergquist, J.~C., Wineland, D.~J., 2006. Spectroscopy of atomic and molecular
  ions using quantum logic vi. In: AIP Conf. Proc. 862. Vol. 862. pp. 305--312.

\bibitem[{Schmidt et~al.(2005)Schmidt, Rosenband, Langer, Itano, Bergquist, and
  Wineland}]{05Schmidt}
Schmidt, P.~O., Rosenband, T., Langer, C., Itano, W.~M., Bergquist, J.~C.,
  Wineland, D.~J., 2005. Spectroscopy using quantum logic. Science 309,
  749--752.

\bibitem[{Schneider et~al.(2010)Schneider, Enderlein, Huber, and
  Schaetz}]{10Schneider}
Schneider, C., Enderlein, M., Huber, T., Schaetz, T., 2010. Optical trapping of
  an ion. Nature Photonics 4, 772.

\bibitem[{Senko et~al.(2012)Senko, Mizrahi, Campbell, Johnson, Conover, and
  Monroe}]{12Senko}
Senko, C., Mizrahi, J., Campbell, W.~C., Johnson, K.~G., Conover, C. W.~S.,
  Monroe, C., 2012. Ultrafast spin-motion entanglement and interferometry with
  a single atom. arXiv:1201.0776.

\bibitem[{S{\o}rensen and M{\o}lmer(2000)}]{00Sorensen1}
S{\o}rensen, A., M{\o}lmer, K., 2000. Entanglement and quantum computation with
  ions in thermal motion. Phys. Rev. A 62, 022311.

\bibitem[{Splatt et~al.(2009)Splatt, Harlander, Brownnutt, Z\"{a}hringer,
  Blatt, and H\"{a}nsel}]{09Splatt}
Splatt, F., Harlander, M., Brownnutt, M., Z\"{a}hringer, F., Blatt, R.,
  H\"{a}nsel, W., 2009. Deterministic reordering of 40 ca + ions in a linear
  segmented paul trap. New Journal of Physics 11~(10), 103008.
\newline\urlprefix\url{http://stacks.iop.org/1367-2630/11/i=10/a=103008}

\bibitem[{Staanum et~al.(2010)Staanum, H{\o}jbjerre, Skyt, Hansen, and
  Drewsen}]{10Staanum}
Staanum, P.~F., H{\o}jbjerre, K., Skyt, P.~S., Hansen, A.~K., Drewsen, M.,
  2010. Rotational laser cooling of vibrationally and translationally cold
  molecular ions. Nature Physics 6, 271--274.

\bibitem[{Steane(1997)}]{97Steane2}
Steane, A.~M., 1997. The ion trap quantum information processor. Appl. Phys. B
  64, 623--642.

\bibitem[{Steane(2003)}]{03Steane}
Steane, A.~M., 2003. Overhead and noise threshold of fault-tolerant quantum
  error correction. Phys. Rev. A 68, 042322.

\bibitem[{Tong et~al.(2010)Tong, Winney, and Willitsch}]{10Tong}
Tong, X., Winney, A.~H., Willitsch, S., Sep 2010. Sympathetic cooling of
  molecular ions in selected rotational and vibrational states produced by
  threshold photoionization. Phys. Rev. Lett. 105, 143001.
\newline\urlprefix\url{http://link.aps.org/doi/10.1103/PhysRevLett.105.143001}

\bibitem[{Turchette et~al.(2000)Turchette, Kielpinski, King, Leibfried,
  Meekhof, Myatt, Rowe, Sackett, Wood, Itano, Monroe, and
  Wineland}]{00Turchette}
Turchette, Q.~A., Kielpinski, D., King, B.~E., Leibfried, D., Meekhof, D.~M.,
  Myatt, C.~J., Rowe, M.~A., Sackett, C.~A., Wood, C.~S., Itano, W.~M., Monroe,
  C., Wineland, D.~J., 2000. Heating of trapped ions from the quantum ground
  state. Phys. Rev. A 61, 063418.

\bibitem[{Vahala et~al.(2009)Vahala, Herrmann, Knunz, Batteiger, Saathoff,
  Haensch, and Udem}]{09Vahala}
Vahala, K., Herrmann, M., Knunz, S., Batteiger, V., Saathoff, G., Haensch,
  T.~W., Udem, T., 2009. A photon laser. Nature Physics 5, 682--686.

\bibitem[{Vogelius et~al.(2002)Vogelius, Madsen, and Drewsen}]{02Vogelius}
Vogelius, I.~S., Madsen, L.~B., Drewsen, M., Oct 2002.
  Blackbody-radiation\char21{}assisted laser cooling of molecular ions. Phys.
  Rev. Lett. 89, 173003.
\newline\urlprefix\url{http://link.aps.org/doi/10.1103/PhysRevLett.89.173003}

\bibitem[{Vogelius et~al.(2006{\natexlab{a}})Vogelius, Madsen, and
  Drewsen}]{06Vogelius}
Vogelius, I.~S., Madsen, L.~B., Drewsen, M., 2006{\natexlab{a}}. Rotational
  cooling of molecular ions through laser-induced coupling to the collective
  modes of a two-ion coulomb crystal. Journal of Physics B: Atomic, Molecular
  and Optical Physics 39~(19), S1267.
\newline\urlprefix\url{http://stacks.iop.org/0953-4075/39/i=19/a=S32}

\bibitem[{Vogelius et~al.(2006{\natexlab{b}})Vogelius, Madson, and
  Drewsen}]{06Vogelius2}
Vogelius, I.~S., Madson, L.~B., Drewsen, M., 2006{\natexlab{b}}. Probabilistic
  state preparation of a single molecular ion by projection measurement. J.
  Phys. B 39, S1259--S1265.

\bibitem[{Wesenberg et~al.(2007)Wesenberg, Epstein, Leibfried, Blakestad,
  Britton, Home, Itano, Jost, Knill, Langer, Ozeri, Seidelin, and
  Wineland}]{07Wesenberg}
Wesenberg, J.~H., Epstein, R.~J., Leibfried, D., Blakestad, R.~B., Britton, J.,
  Home, J.~P., Itano, W.~M., Jost, J.~D., Knill, E., Langer, C., Ozeri, R.,
  Seidelin, S., Wineland, D.~J., Nov 2007. Fluorescence during doppler cooling
  of a single trapped atom. Phys. Rev. A 76, 053416.

\bibitem[{Wineland and Itano(1979)}]{79Wineland}
Wineland, D.~J., Itano, W.~M., 1979. Laser cooling of atoms. Phys. Rev A.
  20~(4), 1521--1539.

\bibitem[{Wineland et~al.(1998)Wineland, Monroe, Itano, Leibfried, King, and
  Meekhof}]{98Wineland2}
Wineland, D.~J., Monroe, C., Itano, W.~M., Leibfried, D., King, B.~E., Meekhof,
  D.~M., 1998. Experimental issues in coherent quantum-state manipulation of
  trapped atomic ions. J. Res. Natl. Inst. Stand. Technol. 103, 259--328.

\bibitem[{W{\"u}bbena et~al.(2012)W{\"u}bbena, Amairi, Mandel, and
  Schmidt}]{12Wubbena}
W{\"u}bbena, J.~B., Amairi, S., Mandel, O., Schmidt, P.~O., 2012. Sympathetic
  cooling of mixed species two-ion crystals for precision spectroscopy.
  arxiv:1202.2730.

\end{thebibliography}

\end{document}